\documentclass[a4paper,11pt]{article}
\usepackage{caption}
\usepackage{jcappub} % for details on the use of the package, please see the JINST-author-manual
\usepackage{lineno}
\usepackage{xcolor}
\usepackage{multirow}
\usepackage{array}
\usepackage{graphics}
\usepackage{hhline}
\usepackage{makecell}
\usepackage{arydshln}
\usepackage{orcidlink}
\usepackage{subcaption}
\usepackage{graphicx}
\usepackage{soul}
\usepackage[capitalise,nameinlink,noabbrev]{cleveref}
\newcommand{\aap}{Astronomy and Astrophysics}
\newcommand{\apjs}{Astrophysical Journal, Supplement}

\newcommand{\apjl}{Astrophysical Journal, Letters}
\newcommand{\effort}{\texttt{Effort.jl}}

\newcommand{\turing}{\texttt{Turing.jl}}
\newcommand{\hmpcinv}{\,h\,{\rm Mpc^{-1}}}
\newcommand{\hinvmpc}{\,h^{-1}{\rm Mpc}}
\newcommand{\hinvMpccubed}{\, h^{-3} \, \text{Mpc}^{3}}
\newcommand{\Hquad}{\hspace{0.5em}} 
\makeatletter
\input{aas_macros.sty}
\let\jnl@style=\relax
\makeatother

\title{\boldmath Alleviating prior dependencies for DESI DR1 clustering fits through reparameterization}

\affiliation{Affiliations are in Sec.~\ref{sec:affiliations}}
\author[1,2,3]{M.~Bonici\orcidlink{0000-0002-8430-126X},}
\author[4,5]{S.~Paradiso\orcidlink{0000-0002-2753-5211},}
\author[6]{G.~McGee\orcidlink{0000-0002-6133-9169},}
\author[7,8]{G.~D’Amico\orcidlink{0000-0002-8183-1214},}
\author[9]{M.~Karamanis\orcidlink{0000-0001-9489-4612},}
\author[2,3]{H.~Zhang\orcidlink{0000-0001-6847-5254},}
\author[1,2,3]{W.~J.~Percival\orcidlink{0000-0002-0644-5727},}
\author[10]{J.~Aguilar,}
\author[11]{S.~Ahlen\orcidlink{0000-0001-6098-7247},}
\author[12,13]{D.~Bianchi\orcidlink{0000-0001-9712-0006},}
\author[14]{D.~Brooks,}
\author[15,16]{F.~J.~Castander\orcidlink{0000-0001-7316-4573},}
\author[10]{T.~Claybaugh,}
\author[17]{A.~de la Macorra\orcidlink{0000-0002-1769-1640},}
\author[18,19]{Biprateep~Dey\orcidlink{0000-0002-5665-7912},}
\author[14]{P.~Doel,}
\author[9,10]{S.~Ferraro\orcidlink{0000-0003-4992-7854},}
\author[20,21]{A.~Font-Ribera\orcidlink{0000-0002-3033-7312},}
\author[22,23]{J.~E.~Forero-Romero\orcidlink{0000-0002-2890-3725},}
\author[15,16,24]{E.~Gaztañaga\orcidlink{0000-0001-9632-0815},}
\author[25]{Satya~{Gontcho A Gontcho}\orcidlink{0000-0003-3142-233X},}
\author[26]{G.~Gutierrez,}
\author[27]{C.~Hahn\orcidlink{0000-0003-1197-0902},}
\author[28,29,30]{K.~Honscheid\orcidlink{0000-0002-6550-2023},}
\author[31]{M.~Ishak\orcidlink{0000-0002-6024-466X},}
\author[32]{R.~Joyce\orcidlink{0000-0003-0201-5241},}
\author[33]{R.~Kehoe,}
\author[10]{T.~Kisner\orcidlink{0000-0003-3510-7134},}
\author[10]{A.~Kremin\orcidlink{0000-0001-6356-7424},}
\author[14]{O.~Lahav\orcidlink{0000-0002-1134-9035},}
\author[30]{C.~Lamman\orcidlink{0000-0002-6731-9329},}
\author[10]{M.~Landriau\orcidlink{0000-0003-1838-8528},}
\author[34]{L.~Le~Guillou\orcidlink{0000-0001-7178-8868},}
\author[21,35]{M.~Manera\orcidlink{0000-0003-4962-8934},}
\author[32]{A.~Meisner\orcidlink{0000-0002-1125-7384},}
\author[20,21]{R.~Miquel,}
\author[36,37]{G.~Niz\orcidlink{0000-0002-1544-8946},}
\author[38]{F.~Prada\orcidlink{0000-0001-7145-8674},}
\author[39]{I.~P\'erez-R\`afols\orcidlink{0000-0001-6979-0125},}
\author[40]{G.~Rossi,}
\author[41,42]{L.~Samushia\orcidlink{0000-0002-1609-5687},}
\author[43]{E.~Sanchez\orcidlink{0000-0002-9646-8198},}
\author[44]{E.~F.~Schlafly\orcidlink{0000-0002-3569-7421},}
\author[10]{D.~Schlegel,}
\author[10]{J.~Silber\orcidlink{0000-0002-3461-0320},}
\author[32]{D.~Sprayberry,}
\author[45]{G.~Tarl\'{e}\orcidlink{0000-0003-1704-0781},}
\author[17]{M.~Vargas-Maga\~na\orcidlink{0000-0003-3841-1836},}
\author[32]{B.~A.~Weaver,}
\author[34]{P.~Zarrouk\orcidlink{0000-0002-7305-9578},}
\author[46]{H.~Zou\orcidlink{0000-0002-6684-3997},}

\emailAdd{mbonici@uwaterloo.ca}

\abstract{Bayesian analyses of the full-shape clustering of Dark Energy Spectroscopic Instrument (DESI) Data Release 1 (DR1) exhibit prior-volume projection effects, whereby weakly constrained nuisance parameters of the Effective Field Theory of Large Scale Structure (EFTofLSS) shift marginalized cosmological posteriors away from the posterior maximum. We reanalyze DESI DR1 power spectrum multipoles using two complementary mitigation strategies: (i) nonlinear orthogonalization to decorrelate nuisance and cosmological parameter priors, and (ii) a fully reparameterization-invariant Jeffreys prior over all EFTofLSS coefficients, evaluated on-the-fly via closed-form Jacobians. Including data from DESI, Big-Bang Nuclesynthesis and a constraint on $n_{\mathrm{s}}$, baseline priors lead to multi-$\sigma$ projection in the Hubble parameter $H_{0}$ and dark energy equation of state parameters $w_{0}$ and $w_{a}$; the Jeffreys prior successfully recenters these posteriors to enclose the maximum a posteriori estimate within the 68\% credible regions, demonstrating clear mitigation of projection effects for these late-time expansion parameters. A hybrid Jeffreys+baseline-Gaussian configuration controls residual over-broad tails in the physical cold dark matter density $\omega_{\mathrm{c}}$ while preserving the volume correction, and is our favoured approach. We compare the credible intervals derived using our methodology to those obtained using Halo Occupation Distribution (HOD)-informed priors and to confidence intervals derived using frequentist profile likelihood analyses, finding agreement in both central values and degeneracy directions in the $w_{0}$--$w_{a}$ plane. This demonstrates that, once projection effects are properly controlled, we can make robust inferences about the late-time cosmological expansion independent of the statistical framework adopted. 

}

\begin{document}
\maketitle
\flushbottom

\section{Introduction}

Cosmology has experienced remarkable advances over recent decades, propelled by increasingly precise observations of Supernovae~\cite{SupernovaSearchTeam:1998fmf, SupernovaCosmologyProject:1998vns}, the Cosmic Microwave Background (CMB)~\cite{Collaboration2020PlanckParameters} and the Large-Scale Structure (LSS) of the Universe~\cite{Dawson2013TheSDSS-III, Dawson2016TheData, Alam2021CompletedObservatory, 2013kids, DES:2016jjg, 2018HSC}. Galaxy surveys and intergalactic medium tracers have provided essential insights into matter distribution~\cite{Wright:2025xka, DES:2026fyc}, dark energy (DE) properties~\cite{DESI:2025zgx}, and neutrino mass constraints~\citep{Lesgourgues:2012uu, Elbers:2025vlz, Chebat:2025kes}. These cosmological investigations rely heavily on statistical analyses of clustering patterns, particularly baryon acoustic oscillations (BAO)~\cite{Eisenstein1998CosmicSurveys} and redshift-space distortions (RSD)~\cite{kaiser1987}, which yield constraints on both the Universe's expansion history and cosmic structure growth. A significant leap forward in this class of measurements is represented by Stage-IV surveys including Euclid~\cite{EuclidCollaboration2024Euclid.Mission}, Roman~\cite{Spergel2015Wide-FieldReport},  LSST~\cite{Ivezic2019LSST:Products}, and DESI.

The Dark Energy Spectroscopic Instrument (DESI) is the first operational Stage-IV galaxy survey \cite{Levi2013The2013,DESICollaboration2016TheDesign,Collaboration2024ValidationInstrument,Collaboration2024TheInstrument}, marking a substantial progression in the field. The eight-year spectroscopic campaign of DESI is mapping 17,000 square degrees of the celestial sphere \cite{DESICollaboration2016TheDesignb, Collaboration2022OverviewInstrument,Silber2023TheDESI,Miller2024TheInstrument,Schlafly2023SurveyInstrument,Guy2023TheInstrument,FiberSystem.Poppett.2024}, targeting five distinct cosmic tracers over a wide redshift range of $0 < z < 4$. These tracers consist of the Bright Galaxy Survey (BGS) \cite{Hahn2023TheValidation}, luminous red galaxies (LRG) \cite{Zhou2023TargetGalaxies}, emission line galaxies (ELG) \cite{Raichoor2023TargetGalaxies}, quasars (QSO) \cite{Chaussidon2023TargetQuasars}, and the Ly$\alpha$ forest \cite{Myers2023TheInstrument}. The survey aims to collect precise redshifts for approximately 63 million galaxies and quasars, thereby providing a highly detailed map of the Universe's large-scale structure.

DESI has publicly released its first large samples of data with Data Release 1 (DR1) \cite{2025arXiv250314745D}.  These data have enabled important cosmological results through detailed galaxy clustering measurements \cite{DESICollaboration2024DESIQuasars,Adame2025DESIOscillations,Adame2025DESIForest,DESICollaboration2024DESIQuasarsb,DESICollaboration2024DESIMeasurements, Ishak2024Modified2024}. A cornerstone of this progress stems from Baryon Acoustic Oscillation (BAO) measurements, where the characteristic oscillatory pattern in the clustering power spectrum serves as a robust standard ruler for probing cosmic expansion \cite{Eisenstein1998CosmicSurveys,Blake2003ProbingRuler,Seo2003ProbingSurveys}.

While BAO analysis provides valuable cosmological constraints, the next frontier lies in Full-Shape (FS) clustering measurements of DESI's tracers. Although this approach demands more sophisticated modeling, it offers substantial advantages by capturing not only cosmic structure growth signals but also encoding information about the primordial power spectrum's amplitude and shape \cite{Peebles1980TheUniverse, Liddle2000CosmologicalStructure, Koyama2016CosmologicalGravity, Joyce2016DarkGravity, Ishak2019TestingCosmology, Alam2021TowardsRequirements, Huterer2023GrowthStructure}. This comprehensive approach promises to unlock deeper cosmological insights from DESI's unprecedented dataset.

The main approach for modeling full-shape galaxy clustering is the cosmological perturbation theory (PT) framework~\cite{Bernardeau2002Large-scaleTheory}, which is the setting in which prior-volume projection effects are most acute. This theory models the non-linear evolution of the matter power spectrum using a series expansion in increasing powers of the overdensity. This framework has been substantially improved by including effective field theory (EFT) techniques, which systematically introduce counterterms to account for small-scale effects \cite{Baumann2012CosmologicalFluid,Carrasco2012TheStructures,Porto2014TheStructures,Perko2016BiasedStructure, Lewandowski2017AnRegime,Ivanov2022EffectiveStructure}. The EFT of Large-Scale Structure (EFTofLSS) is expected to describe every aspect of the long-wavelength dynamics of the universe, including: dark matter~\cite{Baumann:2010tm}, baryons~\cite{Lewandowski:2014rca}, neutrinos~\cite{Senatore:2017hyk, Aviles:2021que}, and time-dependent, smooth, or clustering dark energy~\cite{Lu:2025gki, DAmico:2020tty}. This includes predictions for both density and velocity fields in real and redshift space~\cite{Senatore:2014vja, Ivanov:2018gjr}. The EFT approach addresses non-linear gravitational dynamics and the complex processes of galaxy formation and distribution, making it an essential tool for deriving reliable cosmological constraints from galaxy clustering observations \cite{Colas2020EfficientStructure,dAmico2020TheStructure,Ivanov2020CosmologicalSpectrum,DAmico2021LimitsCode,Niedermann2021NewData,Kumar2022UpdatingGalaxies, Simon2022ConstrainingStructures,Nunes2022NewSpectrum,Philcox2022BOSSMonopole,Zhang2022BOSSStructure,Chen2022ABAO,Lague2022ConstrainingSurveys,Simon2023ConsistencySpectrum,Carrilho2023CosmologyPriors,Schoneberg2023ComparativeRadiation,Smith2023AssessingConstant,Allali2023DarkDatasets,Simon2023CosmologicalAnalysis,Simon2023UpdatedEnergy,DAmico2024LimitsEFTofLSS}. Since galaxies are biased tracers of the underlying matter density field, the EFTofLSS must also incorporate a galaxy bias expansion, which parameterizes the galaxy overdensity as a perturbative series in the local matter fields and their derivatives \cite{Desjacques:2016bnm}. Together with the EFT counterterms and stochastic shot-noise contributions, these bias coefficients form the full nuisance sector of the model.

This theoretical approach, however, presents a significant challenge related to the need to marginalize over multiple nuisance parameters within a Bayesian analysis. These parameters, which control galaxy bias, counterterms, and shot noise effects, frequently show strong degeneracies with the cosmological parameters of primary interest. This creates a complex parameter space and complicates inference. When these degeneracies are paired with broad priors, they can lead to projection effects \cite{Simon2023ConsistencySpectrum}. This phenomenon arises when unconstrained areas of the nuisance parameter space have a disproportionate impact on the marginalized posterior of cosmological parameters, causing systematic deviations of the (commonly reported) expectation values of the 1-dimensional marginalized posteriors from Maximum a Posteriori (MAP) values. The issue becomes especially pronounced in extended cosmological models \cite{Simon2023ConsistencySpectrum,Gomez-Valent2022FastCosmology,Carrilho2023CosmologyPriors, Hadzhiyska2023CosmologyParameters, DAmico2024TheStructure, Maus2025AnBeyond}.

Standard analyses typically rely on uniform and Gaussian priors for most nuisance parameters, with prior widths dictated by perturbativity considerations, and calibrated through extensive simulation-based validation campaigns \cite{Nishimichi:2020tvu, Ivanov2020CosmologicalSpectrum, DAmico2021LimitsCode, Philcox2022BOSSMonopole}. Different EFT codes—including \texttt{velocileptors}, \texttt{PyBird}, and \texttt{FOLPSv}—have been shown to deliver consistent constraints when equivalent prior choices are adopted \cite{Maus2025ASurveys}; in particular, the Lagrangian and Euleria implementations of \texttt{velocileptors} yield nearly identical posteriors when sampling in the same Lagrangian bias basis with a subsequent transformation to the Eulerian basis at the likelihood level \cite{DESICollaboration2024DESIQuasarsb}. Despite this inter-code consistency, the standard approach remains vulnerable to projection effects in cosmological parameter values due to the inherently difficult-to-constrain nature of these nuisance parameters.

Several strategies have been proposed to mitigate these systematic challenges. These include the implementation of simulation-based priors for nuisance parameters \cite{Zhang2024HOD-informedLSS, Ivanov2024Full-shapeBOSS,Ivanov2024Full-shapeAnomaly, 2025arXiv250307270I, DESI:2025wzd}. The rationale for this approach is that projection effects are primarily driven by the poorly constrained parameters introduced by the EFTofLSS; consequently, applying more restrictive priors which are meant to represent information on small-scale physics to these parameters can largely ameliorate the issue.
While this approach has been shown to be effective in reducing projection effects and improving the constraining power of FS analyses, it depends on the modeling of the galaxy-halo connection. The implicit assumption is that clustering can be accurately reproduced by Halo Occupation Distribution (HOD) models, either at the two-point function level \cite{Zhang2024HOD-informedLSS, DESI:2025wzd} or at the field level \cite{Ivanov2024Full-shapeBOSS, Ivanov2024Full-shapeAnomaly}. Therefore, if the HOD model employed cannot fully reproduce the observed clustering, this could result in precise but inaccurate cosmological parameter inference.

For this reason, it is important to pursue alternative methods that do not rely on such assumptions. Other approaches currently explored in the literature include the use of non-uniform integration measures~\cite{Reeves:2025bxc, DAmico:2025zui}, physically motivated reparameterizations~\cite{Tsedrik:2025hmj}, perturbativity priors~\cite{DAmico:2022gki, DAmico:2025zui}, Jeffreys priors~\cite{Hadzhiyska2023CosmologyParameters, Donald-McCann:2023kpx, Zhao:2023ebp}, and the adoption of frequentist inference methods that avoid explicit prior dependencies \cite{Holm2023DecayingLikelihoods,Holm2023BayesianData,Herold2024ProfileEnergy,desi-frequentist}.

This work addresses the challenge of projection effects in the cosmological analysis of DESI Data Release 1 (DR1) clustering measurements by conducting a reanalysis based on reparameterization techniques. We investigate two primary methodological approaches: the orthogonalization of nuisance and cosmological parameters following ~\citep{Paradiso2024ReducingStructure} and the implementation of a reparameterization-invariant measure.  The latter approach employs the Jeffreys prior, which can be viewed as a reparameterization technique since it is constructed from the Fisher information matrix and remains invariant under parameter transformations, naturally accounting for the degeneracies and geometry of the parameter space.

This work provides a comprehensive comparative analysis by contrasting our results with those from complementary reanalyses: a frequentist framework approach~\cite{desi-frequentist} and an HOD-calibrated prior methodology~\cite{DESI:2025wzd}. Each framework operates from distinct foundational assumptions, carrying specific advantages and limitations. This comparison of different frameworks enables us to identify both consistencies and discrepancies across approaches, facilitating a more nuanced and robust interpretation of the results. For the HOD-based comparison, we additionally examine the posterior distribution of EFT parameters obtained through our methodology against the distribution employed in the HOD-based analysis, providing deeper insights into the parameter space behavior across different theoretical frameworks.

This paper is structured as follows. In Sec.~\ref{sec:data} we describe the DESI DR1 full-shape measurements, the construction of the data vectors, window and covariance treatments, and the perturbation-theory modeling and parameterization used to fit the power-spectrum multipoles. In Sec.~\ref{sec:decorrelation} we detail the two mitigation strategies for prior-volume projection—nonlinear orthogonalization of nuisance and cosmological parameters, and a fully reparameterization-invariant Jeffreys prior (including a conservative hybrid with baseline EFT widths)—together with implementation choices. In Sec.~\ref{sec:results} we present the results: we compare the debiasing schemes, report constraints for combinations with external datasets, specifically measurements of the Cosmic Microwave Background (CMB) and Type Ia supernovae (SN), and perform cross-framework tests against both the HOD-informed-prior (HIP) pipeline and a frequentist profile-likelihood analysis; we summarize the main findings and implications, emphasizing the reduced prior dependence and the consistency across Bayesian and frequentist inferences. Finally, we summarize our findings in Sec.~\ref{sec:conclusions}.

\section{Data, modeling, and external datasets}
\label{sec:data}
This section provides an overview of the DESI dataset, the techniques used to measure the power spectrum, and the theoretical model based on the perturbation theory framework employed in this paper. This short description is included for completeness; we refer the interested reader to the cited references for a more comprehensive account.

\subsection{DESI Full-Shape measurements}
\label{sec:fs_meassurements}
%The DESI DR1 provides more than 4.7 million redshift measurements across a redshift interval of $0.1 < z < 2.1$, divided into distinct target categories. These include Bright Galaxy Survey (BGS) galaxies ($0.1 < z < 0.4$), Luminous Red Galaxies (LRGs) partitioned into three redshift bins ($0.4 < z < 0.6$, $0.6 < z < 0.8$, and $0.8 < z < 1.1$), Emission Line Galaxies (ELGs) ($1.1 < z < 1.6$), and Quasars (QSOs) ($0.8 < z < 2.1$). A detailed description of the sample selection and its characteristics is provided in \cite{DESICollaboration2024DESIStatistics}.

To enable a direct comparison with the DR1 Full-Shape DESI analysis \cite{ DESICollaboration2024DESIQuasarsb, DESICollaboration2024DESIMeasurements}, we closely follow its scale-cuts, theoretical modeling, and priors. We use the Feldman-Kaiser-Peacock (FKP) \cite{Feldman1994Power-SpectrumSurveys} estimator, as implemented in the \texttt{pypower} code\footnote{\url{https://github.com/cosmodesi/pypower}}, to extract the power spectrum multipole measurements from the DESI tracers \cite{Yamamoto2006ASurvey, Bianchi2015MeasuringFFTs, Hand2018Nbodykit:Structure}. Weights are assigned to the galaxies to account for the selection function and optimize the measurement of two-point statistics; the codes and weighting scheme are described in detail in \cite{DESICollaboration2024DESIStatistics}[Sec.~8]. The small-scale signal from DESI's fiber assignment process is mitigated by combining the $\theta$-cut method \cite{Pinon2025MitigationEstimators}[Secs.~2.4--2.5] with a rotation of the data vector, window matrix, and covariance, which creates a more diagonal window function \cite{Pinon2025MitigationEstimators}[Sec.~5].
 The covariance matrix is derived from 1000 EZmocks and subsequently rescaled to match the semi-empirical covariance determined from the observed data \cite{Forero-Sanchez2024AnalyticalResults, Rashkovetskyi2025Semi-analyticalData,KP4s8-Alves,KP3s8-Zhao}.
Contributions from various systematic effects are incorporated directly into this covariance, including the prior-weight effect; this contribution was quantified in \cite{Findlay2024ExploringAnalysis}. Although our improved prior-mitigation strategy is expected to reduce this effect, the size of that reduction has not yet been quantified, so we conservatively retain the additional covariance contribution recommended by \cite{Findlay2024ExploringAnalysis}. Consistent with the DR1 DESI analysis, our work focuses on the monopole and quadrupole measurements over the wavenumber range $0.02<k<0.2 \hmpcinv$, with a binning width of $\Delta k=0.005 \hmpcinv$.

\subsection{PT-based modeling}
\label{subsubsec:fs_model}
% \begin{table}
% \centering
% \renewcommand{\arraystretch}{1.1}
%     \begin{tabular}{|l|cccc|}
%     \hhline{|=====|}
%     \textbf{Tracer} & $z_{\rm eff}$ & $\sigma^2_n [\hinvMpccubed]$ & $f_{\rm sat}$ & $\sigma_v[\hinvmpc]$\\
%     \hline
%     \texttt{BGS} & 0.295 & 5723 &  0.15&  5.06\\
%     \texttt{LRG1} & 0.510 & 5082 &  0.15&  6.20\\
%     \texttt{LRG2} & 0.706 & 5229 &  0.15&  6.20\\
%     \texttt{LRG3} & 0.930 & 9574 &  0.15&  6.20\\
%     \texttt{ELG} & 1.317 & 10692 &  0.10&  3.11\\
%     \texttt{QSO} & 1.491 & 47377 &  0.03&  5.68\\
%     \hline    
%     \end{tabular}

% \caption{Physical basis conversion parameters for each DESI DR1 tracer, including the effective redshift $z_{\rm eff}$, shot noise amplitude $\sigma^2_n$ in $\hinvMpccubed$, satellite fraction $f_{\rm sat}$, and characteristic velocity $\sigma_v$ in $\hinvmpc$.}
% \label{tab:config}
% \end{table}

\begin{table}
\centering
\renewcommand{\arraystretch}{1.1}
\begin{tabular}{|l|cccccc|}
\hhline{|=======|}
\textbf{Quantity} & \texttt{BGS} & \texttt{LRG1} & \texttt{LRG2} & \texttt{LRG3} & \texttt{ELG2} & \texttt{QSO} \\
\hline
$z_{\rm eff}$ & 0.295 & 0.510 & 0.706 & 0.919 & 1.317 & 1.491 \\
$\sigma^2_n$ $[\hinvMpccubed]$ & 5723 & 5082 & 5229 & 9574 & 10692 & 47377 \\
$f_{\rm sat}$ & 0.15 & 0.15 & 0.15 & 0.15 & 0.10 & 0.03 \\
$\sigma_v$ $[\hinvmpc]$ & 5.06 & 6.20 & 6.20 & 6.20 & 3.11 & 5.68 \\
\hline
\end{tabular}
\caption{Relevant quantities used for basis conversion for each DESI tracer, listing the effective redshift ($z_{\rm eff}$), Poisson shot noise amplitude ($\sigma^2_n$ in $\hinvMpccubed$), satellite fraction ($f_{\rm sat}$), and characteristic velocity dispersion ($\sigma_v$ in $\hinvmpc$).}
\label{tab:config}
\end{table}

To model the measurements, we employ a perturbation theory framework that directly fits the multipoles of the full-shape power spectrum. This analysis uses the one-loop Eulerian Perturbation Theory (EPT) model for the redshift-space galaxy power spectrum, as provided by the \texttt{velocileptors} code\footnote{\url{https://github.com/sfschen/velocileptors}}~\cite{Chen2020ConsistentTheory, Chen2021Redshift-spaceTheory}; this modeling framework includes counterterms to account for the effects of small-scale physics, such as galaxy formation processes and fingers-of-God effects, and incorporates stochastic terms to model contributions from shot noise. To maximize computational performance and ensure compatibility with gradient‑based sampling—crucial when decorrelating cosmological and EFT parameters—we employ the \effort{}\footnote{\url{https://github.com/CosmologicalEmulators/Effort.jl}} emulator~\cite{Bonici2025Effort:Universe}, coupled with \turing{}\footnote{\url{https://turinglang.org/}} to access gradient‑based samplers~\cite{Maus2025ASurveys}.

To handle long-wavelength displacements, which have a large effect on BAO scales, we apply infrared resummation to the power spectrum \cite{Senatore2014RedshiftStructures, Senatore2015TheStructures, Lewandowski2020AnPeak}. The choice of a specific perturbation theory code is not expected to influence the final cosmological results, as demonstrated in \cite{Maus2025ASurveys}. A more detailed description of the model can be found in \cite{Chen2020ConsistentTheory, Chen2021Redshift-spaceTheory}.

We choose our model parameterization to align with the parameter degeneracies observed in the data, which helps match the power spectrum multipoles and mitigate prior volume biases (see Appendix~B.2 of \cite{Maus2025AnBeyond} for details). The set of variable parameters is:
\begin{align}
    \{b_{\rm 1p}, b_{\rm 2p}, b_{\rm sp}, b_{\rm 3p},\alpha_{\rm 0p},\alpha_{\rm 2p},\alpha_{\rm 4p},{\rm SN}_{\rm 0p},{\rm SN}_{\rm 2p},{\rm SN}_{\rm 4p}\}\,.
    \label{eq:basis}
\end{align}
This set includes the galaxy bias parameters ($b_{\rm 1p}$, $b_{\rm 2p}$, $b_{\rm sp}$, $b_{\rm 3p}$), the counterterm parameters ($\alpha_{\rm 0p}$, $\alpha_{\rm 2p}$, $\alpha_{\rm 4p}$), and the stochastic parameters (${\rm SN}_{\rm 0p}$, ${\rm SN}_{\rm 2p}$, ${\rm SN}_{\rm 4p}$). The subscript 'p' indicates that these quantities are defined in the physical basis.

We fix the bias parameter $b_{\rm 3p}$ to zero, as it is predicted to be small and is degenerate with other nuisance parameters \cite{Maus2025AnBeyond,2014PhRvD..90l3522S,2018JCAP...09..008L}. Furthermore, we also set $\alpha_{\rm 4p}$ to zero because it is completely degenerate with $\alpha_{\rm 0p}$ and $\alpha_{\rm 2p}$ when not including the hexadecapole, and we set ${\rm SN}_{\rm 4p}$ to zero since it is expected to be very small~\cite{DESICollaboration2024DESIMeasurements}.

This physical basis is converted to the Eulerian basis via the following relations:
\begin{equation}
\begin{aligned}
&b_{\rm 1E} = \frac{b_{\rm 1p}}{\sigma_8}, \Hquad
b_{\rm 2E} = \frac{b_{\rm 2p}}{\sigma_8^2} + \frac{8}{21}\left(\frac{b_{\rm 1p}}{\sigma_8} - 1\right), \Hquad
b_{\rm sE} = \frac{b_{\rm sp}}{\sigma_8^2} - \frac{2}{7}\left(\frac{b_{\rm 1p}}{\sigma_8} - 1\right), \Hquad
b_{\rm 3E} = \frac{3b_{\rm 3p}}{\sigma_8^3} + \frac{b_{\rm 1p}}{\sigma_8} - 1, \\
&\alpha_{\rm 0E} = \left(\frac{b_{\rm 1p}}{\sigma_8}\right)^2 \alpha_{\rm 0p}, \Hquad
\alpha_{\rm 2E} = f\,\frac{b_{\rm 1p}}{\sigma_8}(\alpha_{\rm 0p}+\alpha_{\rm 2p}), \Hquad
\alpha_{\rm 4E} = f\left(f\,\alpha_{\rm 2p} + \frac{b_{\rm 1p}}{\sigma_8}\alpha_{\rm 4p}\right), \Hquad
\alpha_{\rm 6E} = f^2\,\alpha_{\rm 4p}, \\
&{\rm SN}_{\rm 0E} = {\rm SN}_{\rm 0p}\, \sigma^2_n,\quad
{\rm SN}_{\rm 2E} = {\rm SN}_{\rm 2p}\, \sigma^2_n\, f_{\rm sat}\, \sigma_v^2, \quad
{\rm SN}_{\rm 4E} = {\rm SN}_{\rm 4p}\, \sigma^2_n\, f_{\rm sat}\, \sigma_v^4.
\end{aligned}
\label{eqn:EPT_conversion}
\end{equation}
Parameters with the subscript ``E'' are defined in the Eulerian basis. In these equations, $\sigma_8$ and $f$ are the amplitude of mass fluctuations and the growth factor, respectively, evaluated at the effective redshift of the tracer. The term $\sigma_n^2$ represents the Poissonian shot noise for a tracer, while $f_{\rm sat}$ and $\sigma_v$ correspond to the expected satellite fraction and velocity dispersion for the tracer. 
The specific calibrated values for each tracer are summarized in Fig.~\ref{tab:config}. The determination of $z_{\rm eff}$ and $\sigma_n^2$ is described in \cite{DESICollaboration2024DESIStatistics}, whereas the calibration of $f_{\rm sat}$ and $\sigma_v$ follows the procedure detailed in \cite{Maus2025AnBeyond}.

This transformed set of parameters,
\begin{align}
\{b_{\rm 1E}, b_{\rm 2E}, b_{\rm sE}, b_{\rm 3E}, \alpha_{\rm 0E}, \alpha_{\rm 2E}, \alpha_{\rm 4E}, \alpha_{\rm 6E}, {\rm SN}_{\rm 0E}, {\rm SN}_{\rm 2E}, {\rm SN}_{\rm 4E}\}
\end{align}
is then passed to \effort{} to compute the theoretical predictions. The priors adopted in the DESI DR1 full-shape analysis are summarized in Table~\ref{tab:priors}; throughout this work, we refer to this choice as the baseline prior.

\begin{table}[h]
\centering
\begin{tabular}{|l|c|}
\hline
parameter & prior \\
\hline
$b_{1\rm{p}}$ & $\mathcal{U}[0,3]$ \\
$b_{2\rm{p}}$ & $\mathcal{N}[0, 5^2]$ \\
$b_{\rm sp}$ & $\mathcal{N}[0, 5^2]$ \\
$\alpha_{\rm 0p}$ & $\mathcal{N}[0, 12.5^2]$ \\
$\alpha_{\rm 2p}$ & $\mathcal{N}[0, 12.5^2]$ \\
$\text{SN}_{\rm 0p}$ & $\propto \mathcal{N}[0, 2^2]$ \\
$\text{SN}_{\rm 2p}$ & $\propto \mathcal{N}[0, 5^2]$ \\
\hline
\end{tabular}
\caption{Nuisance parameters and priors for galaxy power spectrum modeling. Here, $\mathcal{U}$ refers to a uniform prior in the range given, whilst $\mathcal{N}(x, \sigma^2)$ refers to the Gaussian normal distribution with mean $x$ and standard deviation $\sigma$. The parameters $b_{1\rm{p}}$, $b_{2\rm{p}}$, and $b_{\rm sp}$ are bias parameters, $\alpha_{\rm 0p}$ and $\alpha_{\rm 2p}$ are counterterm parameters, and $\text{SN}_{\rm 0p}$ and $\text{SN}_{\rm 2p}$ are stochastic noise parameters. The constant of proportionality in front of the $\text{SN}_{\rm 0p}$ and $\text{SN}_{\rm 2p}$ priors indicates that these priors as written are further scaled with corresponding physically motivated terms; see text for details.}
\label{tab:priors}
\end{table}

\subsection{External datasets and cosmological parameters}

When CMB data are not included in the likelihood, we add an independent measurement of the physical baryon density, $\omega_{\rm b}$, from Big Bang Nucleosynthesis (BBN) and a weak constraint on the spectral index, $n_{\rm s}$, with $10\times$ the width of the Planck 2018 result~\cite{Collaboration2020PlanckParameters}, denoted $n_{\rm s10}$.

For the cosmological sector, we consider the $w_0$--$w_a$CDM model with parameter basis
\[
\{ \ln(10^{10}A_{\rm s}), n_{\rm s}, H_0, \omega_{\rm b}, \omega_{\rm c}, w_0, w_a \}.
\]
Here, \(\ln(10^{10}A_{\rm s})\) and \(n_{\rm s}\) determine the amplitude and tilt of the primordial power spectrum, \(H_0\) is the Hubble constant, \(\omega_{\rm b}\) and \(\omega_{\rm c}\) are the physical baryon and cold-dark-matter densities, and \(w_0\) and \(w_a\) describe the time-dependent dark-energy equation of state~\cite{Chevallier:2000qy,Linder:2002et}. Uniform priors are applied to the cosmological parameters, except for \(n_{\rm s}\) and \(\omega_{\rm b}\) when CMB data are excluded, for which we instead adopt Gaussian priors, \(n_{\rm s} \sim \mathcal{N}(0.9649,\,0.042^2)\) and \(\omega_{\rm b} \sim \mathcal{N}(0.02218,\,0.00055^2)\). The ranges of the uniform priors correspond to the boundaries chosen for our emulators and are listed in Table~\ref{tab:emulator_ranges}.

\begin{table}
\centering

\renewcommand{\arraystretch}{1.1}
    \begin{tabular}{|l|c|}
    \hhline{|=|=|}
    \textbf{Parameter} & \textbf{Emulator Range} \\
    \hline
    $\ln{(10^{10}A_{\rm{s}})}$ & [2., 3.5]\\
    $n_{\rm{s}}$ & [0.85, 1.1] \\
    $H_0$ & [50, 100]\\
    $\omega_{\rm{b}}$ & [0.02, 0.025]\\
    $\omega_{\rm{c}}$ & [0.08, 0.16]\\
    $w_0$ & [-3, 0.5] \\
    $w_a$ & [-3, 2] \\
    \hline
    \end{tabular}

\caption{Emulator ranges for the cosmological parameters.}
\label{tab:emulator_ranges}
\end{table}

\begin{table}
\centering
\resizebox{\columnwidth}{!}{
    \small 
\renewcommand{\arraystretch}{1.1}
    \begin{tabular}{|l|ll|}
    \hhline{|===|}
    \textbf{Name} & \textbf{Description} & \textbf{Ref}\\
    \hline
    DESI       & Combined DESI DR1 FS$+$BAO likelihood & \cite{DESICollaboration2024DESIQuasarsb, DESICollaboration2024DESIMeasurements} \\
    \hline
    CMB        & Planck ``lite" CMB likelihood &\cite{Prince_2019}\\
    \hline
    DESY5 & Type Ia supernova likelihood from DES Year 5 compilation&\cite{DESY5_2024}\\
    \hline
    BBN        & Independent measurement on $\omega_{\rm{b}}$ from Big Bang Nucleosynthesis, $\omega_{\rm{b}}\sim\mathcal{N}(0.02218,0.00055^2)$&\cite{Schoneberg2024TheUpdate}\\
    $n_{\mathrm{s10}}$ & Weak constraint on $n_{\rm{s}}$ with width 10 times wider than \textit{Planck}, $n_{\rm{s}}\sim\mathcal{N}(0.9649,0.042^2)$&\cite{Collaboration2020PlanckParameters}\\
    \hline
    \end{tabular}
}
\caption{Summary of datasets used in this analysis. The first column lists the shorthand notation for each likelihood, followed by a brief description and relevant references.}
\label{tab:likelihoods}
\end{table}

To accelerate our analysis, we employ surrogate models: specifically, \texttt{Effort.jl}~\cite{Bonici2025Effort:Universe} to emulate the FS power-spectrum multipoles and \texttt{Capse.jl}~\cite{Bonici:2023xjk} to model the CMB primary-anisotropy power spectrum. This approach offers two main advantages: it significantly speeds up theoretical calculations and, because these codes are differentiable~\cite{Blondel_2024}, it enables the use of gradient-based methods. In the context of cosmological summary statistics, recent studies have demonstrated the promise of such techniques for further accelerating analysis pipelines~\cite{Bonici:2022xlo, Campagne_2023, Piras_2023, Ruiz-Zapatero_2024, Nygaard_2023, Balkenhol_2024}.

For the CMB likelihood, we adopt the compressed 2018 Planck likelihood\footnote{\href{https://github.com/JuliaCosmologicalLikelihoods/PlanckLite.jl}{\texttt{PlanckLite.jl}}} developed in~\cite{Prince_2019}. By marginalizing over CMB-specific nuisance parameters, this methodology reduces computational complexity. Comparisons within the CMB community show close agreement between marginalized and full likelihoods~\cite{Prince_2019, Planck_2018_CMBlikelihood, Balkenhol_2025, ACT_2025}. Finally, we use the Type Ia supernova likelihood from DES Year 5, implemented in \texttt{Julia}\footnote{\href{https://github.com/JuliaCosmologicalLikelihoods/SNIaLikelihoods.jl}{\texttt{SNIaLikelihoods.jl}}} to be compatible with the other employed likelihoods. Table~\ref{tab:likelihoods} summarizes the datasets included in this paper.

\section{Methodology: Strategies for mitigating projection effects}
\label{sec:decorrelation}

In this section, we describe the reparameterization method that decorrelates the dependence of the posterior to nuisance and cosmological parameters to minimize the sensitivity to the priors placed on nuisance parameters. We also consider adopting the Jeffreys prior, a reparameterization-invariant volume measure, and describe its practical evaluation via the Fisher information (including a conservative hybrid that combines Jeffreys with the baseline Gaussian envelope). Both of these strategies — nonlinear orthogonalization and Jeffreys-based volume correction — have the potential to suppress projection effects.

\subsection{Orthogonal Reparameterization}
\label{sec:orthogonalization}

The orthogonal reparameterization technique is designed to systematically transform the parameter space to decorrelate the nuisance parameters prior and cosmological parameter posterior. Let the full parameter vector be denoted by $\boldsymbol{\theta} = [\mathbf{C}, \mathbf{N}]$, where $\mathbf{C}$ represents the cosmological parameters of interest and $\mathbf{N}$ is the set of nuisance parameters. The posterior distribution, $\pi(\mathbf{C}, \mathbf{N} | \mathbf{y})$, derived from the data $\mathbf{y}$, typically shows coupling between $\mathbf{C}$ and $\mathbf{N}$. Our approach is to find a transformation of the parameter space such that $\mathbf{C}$ and a newly defined set of parameters $\mathbf{N}'$ are rendered approximately decorrelated. This is achieved by first learning the conditional expectation \(f_j(\mathbf{C})=\mathbb{E}[N_j\mid\mathbf{C}]\) from the initial posterior and then transforming \(N'_j=N_j-f_j(\mathbf{C})\) so that \(\mathbf{C}\) and \(\mathbf{N}'\) are approximately uncorrelated, thereby reducing projection effects and stabilising inference~\cite{Paradiso2024ReducingStructure}. In the originally proposed implementation each \(f_j\) was estimated with penalised spline bases within a Generalized Additive Model (GAM) using \texttt{pyGAM}\footnote{\url{https://github.com/dswah/pyGAM}}, selecting smoothing parameters via restricted maximum likelihood, and the learned mapping was evaluated inside the \texttt{Julia}~\cite{Bezanson:2014pyv} inference stack so the inverse transformation \(N_j=N'_j+f_j(\mathbf{C})\) could be applied during likelihood evaluation with \turing{} and \effort{}. However, GAMs struggle to capture the pronounced multivariate, highly non-linear degeneracies among cosmological parameters unless tensor–product smooths are introduced, which rapidly increases the number of basis functions and computational cost. Furthermore, spline extrapolation is numerically brittle when the transformation is evaluated outside the typical set of the training posterior, yielding unstable behavior in high dimensions. We therefore replaced the GAM with a fully connected neural network (NN) that naturally encodes multivariate non-linear interactions, scales gracefully with dimension, extrapolates more smoothly, and in practice delivers faster and more reliable convergence.
Concretely, we worked as follows:
\begin{itemize}
    \item Run an initial analysis with standard priors to obtain posterior samples $S$ \(\{(\mathbf{C}_s,\mathbf{N}_s)\}_{s=1}^S\).
    \item Train a NN regressor \(f(\mathbf{C}) \approx \mathbb{E}[\mathbf{N}\mid \mathbf{C}]\) on these samples, to learn the degeneracies among cosmological and nuisance parameters.
    \item Orthogonalise the nuisance parameters via \(\mathbf{N}'=\mathbf{N}-f(\mathbf{C})\) and refit the model in the transformed space \([\mathbf{C},\mathbf{N}']\) using physically motivated priors on \(\mathbf{C}\) and conservative priors on \(\mathbf{N}'\).
    \item Following \cite{Paradiso2024ReducingStructure}, assign zero-mean, unit-variance Gaussian priors in the reparameterised basis, draw prior samples, map them back to the original basis to measure induced variances, and rescale the reparameterised priors so that the mapped samples exhibit the desired variance in the original basis.
    \item Iterate the entire reparameterisation using the new posterior samples until the contours cease to change appreciably, indicating convergence.
\end{itemize}

We emphasise that while the specific regressor employed here changed compared to the one used in \cite{Paradiso2024ReducingStructure}, the overall algorithmic framework remained unchanged throughout our analysis.

\subsection{Jeffreys prior}
\label{sec:jeffreys_prior}
A common class of priors employed when little or no external information is available is represented by the Jeffreys prior~\cite{jeffreys}.
For a parameter vector $\boldsymbol{\theta}$ the Jeffreys prior reads
\begin{equation}
    p_{\mathrm{J}}(\boldsymbol{\theta}) \;\propto\;
    \sqrt{\det \bigl[\mathbf{I}(\boldsymbol{\theta})\bigr]},
\end{equation}
where $\mathbf{I}(\boldsymbol{\theta})$ is the Fisher--information matrix.  Because the Fisher information transforms as a tensor, $p_{\mathrm{J}}$ is \emph{invariant under smooth re-parameterisations}, thereby avoiding hidden preferences that stem from a particular coordinate choice.

The Jeffreys prior is frequently characterized as the \emph{least informative} prior in one dimension, as it is the prior that maximizes the Kullback--Leibler divergence between the prior itself and the posterior~\cite{bernardo1979}.  It is however important to notice that the Jeffreys prior is not free from pathologies: in higher-dimensional spaces the Jeffreys prior can become impractical because its Fisher-information determinant can be numerically unstable, yielding an improper or highly peaked density that hampers numerical sampling and can re-introduce unwanted prior biases despite its coordinate-invariance. A further practical limitation is that, while the Jeffreys prior does not depend on any specific noise realisation, being constructed from the \emph{expected} Fisher information, i.e.\ from the data covariance matrix rather than the observed data vector, it does change whenever the covariance structure changes, for instance when combining with post-reconstruction BAO data or weak lensing. This makes it less portable across dataset combinations compared to theory-level reparameterisations, such as the AP-amplitude scaling of~\cite{Tsedrik:2025hmj}, which constitute fixed coordinate transformations of the model and are therefore independent of the covariance structure.

The Jeffreys prior has previously been used in analyses using EFTofLSS models~\cite{Donald-McCann:2023kpx, Zhao:2023ebp}. In these analyses, the authors mainly had to deal with two issues. First, computing the determinant of the Fisher–information matrix, $\det\mathbf{I}(\boldsymbol{\theta})$, for dozens of correlated parameters can be prohibitively expensive; hence the Jeffreys prescription in previous work was limited to the linear subset of parameters, for which the implementation is straightforward. This constitutes the so called \textit{partial Jeffreys prior} that we extend in this work to include the nonlinear parameters as well. Second, because Jeffreys priors place no explicit limits on the domain of validity, they allow parameters to wander into regions where perturbation theory breaks down, so the authors placed additional constraints to reduce the allowed range for EFT parameters.

Here we extend the prior to all EFT nuisance parameters by evaluating the Fisher matrix at every MCMC step and updating the Jeffreys contribution on-the-fly, as follows. 
For a multivariate-Gaussian likelihood with fixed data covariance
$\mathbf{\Sigma}_{\mathrm{d}}$, the Fisher matrix with respect to the nuisance
sector $\mathbf{N}$ can be written solely in terms of first derivatives,
\begin{equation}
    I_{ij}(\mathbf{N},\mathbf{C}) \;=\;
    \bigl[\mathbf{J}^{\mathsf{T}}\,
          \mathbf{\Sigma}_{\mathrm{d}}^{-1}\,
          \mathbf{J}\bigr]_{ij},
    \qquad
    J_{\ell i} \equiv
    \frac{\partial P_\ell(k;\mathbf{C},\mathbf{N})}{\partial N_i},
\end{equation}
with $P_\ell$ the model power-spectrum multipoles ($\ell=0,2$ in this work).
All entries of the Jacobian $\mathbf{J}$ admit closed-form expressions.

We implemented analytical Jacobians in \texttt{Effort.jl} to enable efficient Jeffreys prior evaluation. Since \texttt{Effort.jl} employs Hamiltonian MonteCarlo sampling~\cite{betancourt2017conceptual}—which already requires automatic differentiation of the joint posterior—incorporating the Jeffreys prior through autodiff would demand second-order derivatives, a computationally expensive operation. To circumvent this challenge while preserving full differentiability, we derived analytical expressions for $\mathbf{J} = \partial \mathbf{P}_{\ell} / \partial \boldsymbol{\theta}_{\mathrm{EFT}}$ and verified them against both computer algebra systems and autodiff calculations, confirming agreement to floating-point precision. Because forming $\mathbf{J}$ and its low-dimensional determinant is inexpensive compared to computing the theory spectra, the Jeffreys factor can be updated at each Hamiltonian step without hampering sampler efficiency. This delivers a fully reparameterization-invariant prior for all EFT coefficients, removing the residual projection effects that persist when only linear terms receive Jeffreys weighting.

Regarding the latter, we consider two scenarios: one when we include only the Jeffreys prior and one when we include the baseline prior on EFT parameters as well. A further analysis comparison using the partial Jeffreys prior is presented in Appendix~\ref{app:full}.

\section{Results}
\label{sec:results}
%We begin by contrasting the marginalised posterior constraints obtained with our two Bayesian mitigation strategies against the maximum-a-posteriori (MAP) estimates of the baseline analysis.  

We now contrast results from the original baseline prior (BLP), the non-linear orthogonalisation scheme, the fully volume-correcting Jeffreys prior, and the Jeffreys prior supplemented by the original baseline priors (J+BLP) on EFT coefficients.

\subsection{Comparison of debiasing methods}

\begin{figure}[htbp]         % “float” so LaTeX can move it
  \centering
  % width=.9\linewidth scales the graphic to 90 % of the text width
  \includegraphics[width=\linewidth]{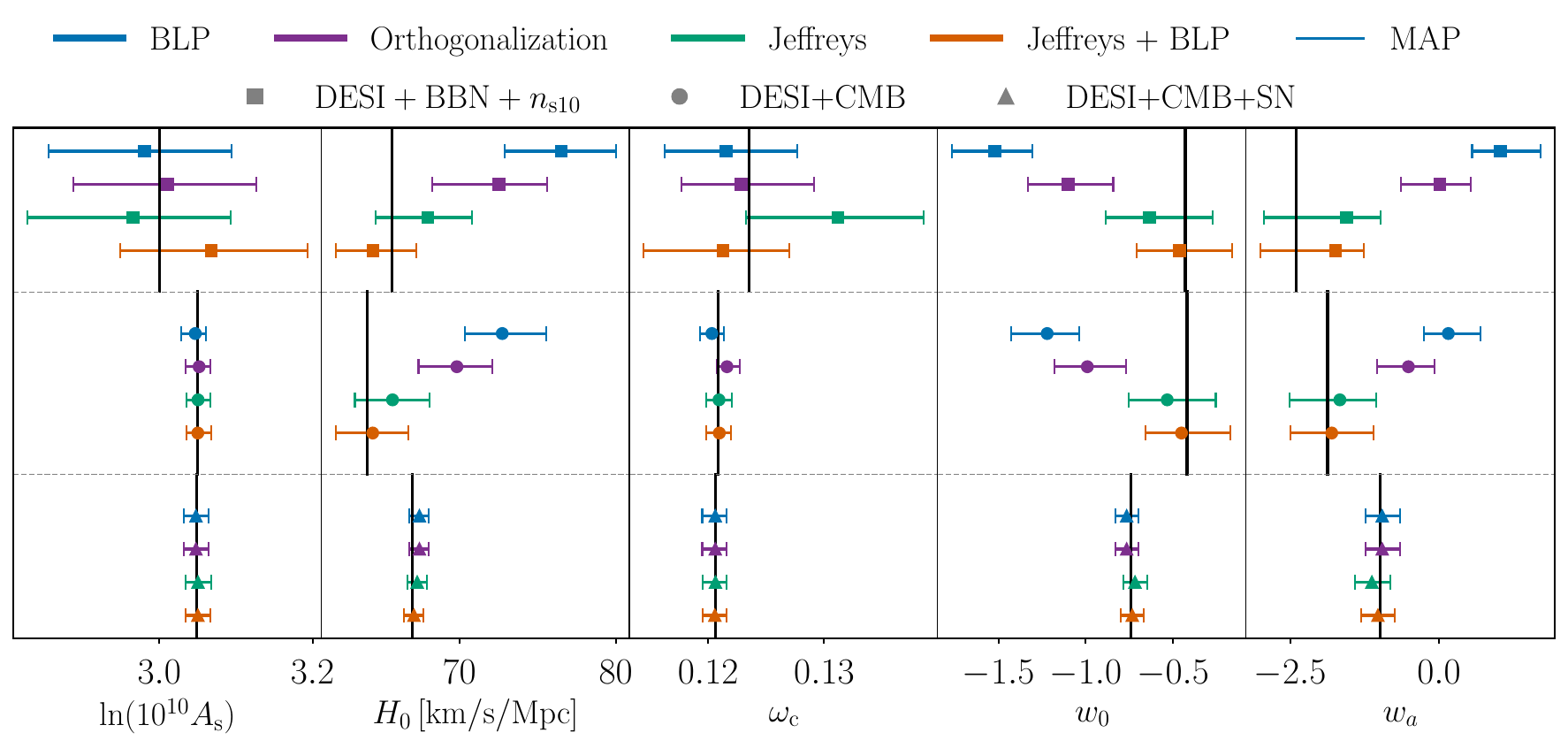}
  \caption{Comparison of projection-mitigation schemes across dataset combinations. For each cosmological parameter, we show the MAP (vertical gray line) and 68\% marginalized credible intervals from four analysis approaches: BLP (blue), orthogonalization (purple), Jeffreys prior (green), and hybrid Jeffreys+BLP (orange). Results are presented for three dataset combinations: $\textsc{DESI}+\textsc{BBN}+n_{\mathrm{s},10}$ (squares), $\textsc{DESI}+\textsc{CMB}$ (circles), and $\textsc{DESI}+\textsc{CMB}+\textsc{SN}$ (triangles). In the weakly anchored $\textsc{DESI}+\textsc{BBN}+n_{\mathrm{s},10}$ configuration, BLP exhibits significant projection effects with posteriors displaced several standard deviations from the MAP, particularly for $H_{0}$, $w_{0}$, and $w_{a}$. Orthogonalization partially mitigates these offsets, while both Jeffreys variants substantially recenter the posteriors to enclose the MAP. With stronger external constraints ($\textsc{CMB}$ and $\textsc{SN}$), all four approaches converge, indicating that projection effects are effectively eliminated in richer data combinations.}
  \label{fig:cosmo_whiskers}
\end{figure}

Fig.~\ref{fig:cosmo_whiskers} compares the three projection--mitigation schemes—orthogonalisation, the Jeffreys prior, and the hybrid Jeffreys+baseline--Gaussian—across all dataset combinations considered in this work, showing for each cosmological parameter the MAP (vertical lines) together with the corresponding 68\% credible intervals. We begin with the \textsc{DESI,+ BBN,+ $n_{\mathrm{s},10}$} case, where the weak external anchoring makes projection most severe: the baseline  analysis (BLP, blue) yields posteriors for $H_0$ and $(w_0,w_a)$ that lie several standard deviations from the MAP; orthogonalisation (purple) mitigates but does not eliminate this mismatch, typically moving the means of $H_0$, $w_0$, and $w_a$ inward by $\sim1\sigma$; the Jeffreys prior (green) further reduces these offsets; and the hybrid Jeffreys+baseline--Gaussian variant (orange) achieves the closest agreement while simultaneously controlling overly broad EFT directions. In this DESI-only configuration, the pure Jeffreys fit can leave $\omega_\mathrm{c}$ somewhat over-broadened and shifted, whereas the hybrid choice tames this tail without degrading the gains in $H_0$, $w_0$, and $w_a$, motivating its use as our fiducial configuration. 

Turning to the \textsc{DESI\,+\,CMB} combination, we find that the baseline Gaussian analysis is already less afflicted by projection, yet non–negligible offsets persist for $H_{0}$ as well as for the DE parameters.  Orthogonalisation again nudges the posteriors in the right direction but still leaves the MAP outside the 68\% band for all three of these quantities.

Both implementations of the Jeffreys prior succeed where orthogonalisation falls short: their error bars comfortably envelop the MAP for $H_{0}$, $w_{0}$ and $w_{a}$, indicating a reduction in projection effects.  Moreover, the two Jeffreys variants now agree at the sub-\(0.4\sigma\) level across every cosmological parameter, showing that once the CMB prior has tightened the DESI contours, the additional Gaussian envelope on the EFT sector has only a minute impact on the cosmology. 

Finally, with Type Ia supernovae added to the \textsc{DESI\,+\,CMB} data vector, there is a significant reduction of prior-volume artefacts.  Consequently the baseline Gaussian analysis, the orthogonalised fit, and both flavours of the Jeffreys prior all recover consistent posteriors for $w_{0}$ and $w_{a}$, and the MAP is captured by every 68\% interval displayed in Figure~\ref{fig:cosmo_whiskers}.  The same holds for $H_{0}$ and $\omega_\mathrm{c}$: any residual offsets are well below the $0.3\sigma$ level, confirming that projection effects are significantly reduced once the supernova information is included.

In summary, our side‐by‐side comparison across progressively richer data sets shows that prior–volume projection is severe for the \textsc{DESI\,+\,BBN\,+\,$n_\mathrm{s}$} combination, partially alleviated by orthogonalisation procedure and mostly removed by the full Jeffreys prior; that the same hierarchy of performance persists, though at diminished amplitude, once CMB information is added; and that all three debiasing schemes converge when Type Ia supernovae tighten the DE sector.  Because the Jeffreys prior supplemented by the original EFT widths consistently contains the MAP for \emph{every} cosmological parameter in all data configurations—while the pure Jeffreys version shows a residual tail in $\omega_\mathrm{c}$ for DESI–only—we adopt the Jeffreys\,+\,BLP prescription as our fiducial choice for the remainder of this work.

\subsection{Cosmological parameter posterior}

Having established the Jeffreys–augmented baseline prior (BLP+J) as the fiducial configuration for our full–shape analysis — on account of its ability to remove the bulk of prior–volume projection — we now compare its cosmological constraints against those obtained with the standard baseline Gaussian prior (BLP) and with the HOD–informed–prior (HIP) framework \cite{DESICollaboration2024DESIMeasurements,DESICollaboration2024DESIQuasarsb,DESI:2025wzd}. The comparison to BLP clarifies how marginalized posteriors shift when projection is mitigated, whereas the comparison to HIP serves as a validation against a complementary, simulation–calibrated approach that injects galaxy–halo modeling information; together these tests assess whether our nuisance–agnostic strategy introduces tensions in the recovered cosmology or, conversely, corroborates the HIP conclusions under a distinct set of assumptions \cite{DESICollaboration2024DESIMeasurements,DESI:2025wzd}. Additionally, tests in Appendix~\ref{app:test} show the recovery of the input parameters when using our fiducial approach on noiseless datavectors.

To ensure an apples–to–apples assessment, we enforce methodological parity across pipelines: identical data vectors, $k$–ranges, multipole content, survey window treatment, and covariance prescriptions; the same EFTofLSS model as implemented in \effort{} with the same nuisance basis and Eulerian–to–physical mappings; identical cosmological parameterizations and external anchors (BBN and a loose $n_\mathrm{s}$ when CMB is absent). Under this protocol the sole procedural changes are the nuisance–sector measures: Gaussian widths for BLP, Jeffreys weighting with the baseline envelope for BLP+J, and simulation–calibrated densities for the HIP framework \cite{DESICollaboration2024DESIMeasurements,DESICollaboration2024DESIQuasarsb,DESI:2025wzd}.

Conceptually, BLP+J and HIP address the same failure mode—degeneracy–driven leakage from weakly constrained EFT parameters—via different philosophies: BLP+J enforces a reparameterization–invariant measure that removes prior–volume geometry without external astrophysical assumptions, whereas HIP restricts the nuisance parameter space to galaxy–halo–consistent regions learned from HOD–based mock suites. Agreement between their cosmological posteriors is therefore a nontrivial cross–check; conversely, any systematic offsets (as well as reduction in error bars) would isolate which nuisance directions are most sensitive to galaxy–halo modeling and warrant further scrutiny in forthcoming analyses. This comparison is crucial precisely because the methods encode different assumptions: a statistical, measure–based debiasing in BLP+J versus a physically motivated, HOD–calibrated constraint in HIP. If the HOD adopted within HIP were overly restrictive, it could propagate modeling bias into the cosmological sector; consequently, concordance with BLP+J strengthens the HIP results by indicating that its physical priors are not unduly constraining, while any residual discrepancies would flag specific nuisance directions that require relaxation, alternative calibrations, or additional robustness checks.

In the following, we present this comparison in detail. We begin with DESI\,+\,BBN\,+\,$n_\mathrm{s}$, where projection is most acute and debiasing strategies show the largest impact, then escalate to DESI\,+\,CMB and DESI\,+\,CMB\,+\,SN to test the persistence of any differences as external anchors tighten. We close by contrasting the EFT coefficient posteriors under BLP+J with the HIP prior volumes, highlighting directions where the data either validate or challenge the simulation–calibrated expectations \cite{DESICollaboration2024DESIMeasurements,DESI:2025wzd}. 

\begin{figure}[htbp]         % “float” so LaTeX can move it
  \centering
  % width=.9\linewidth scales the graphic to 90 % of the text width
  \includegraphics[width=\linewidth]{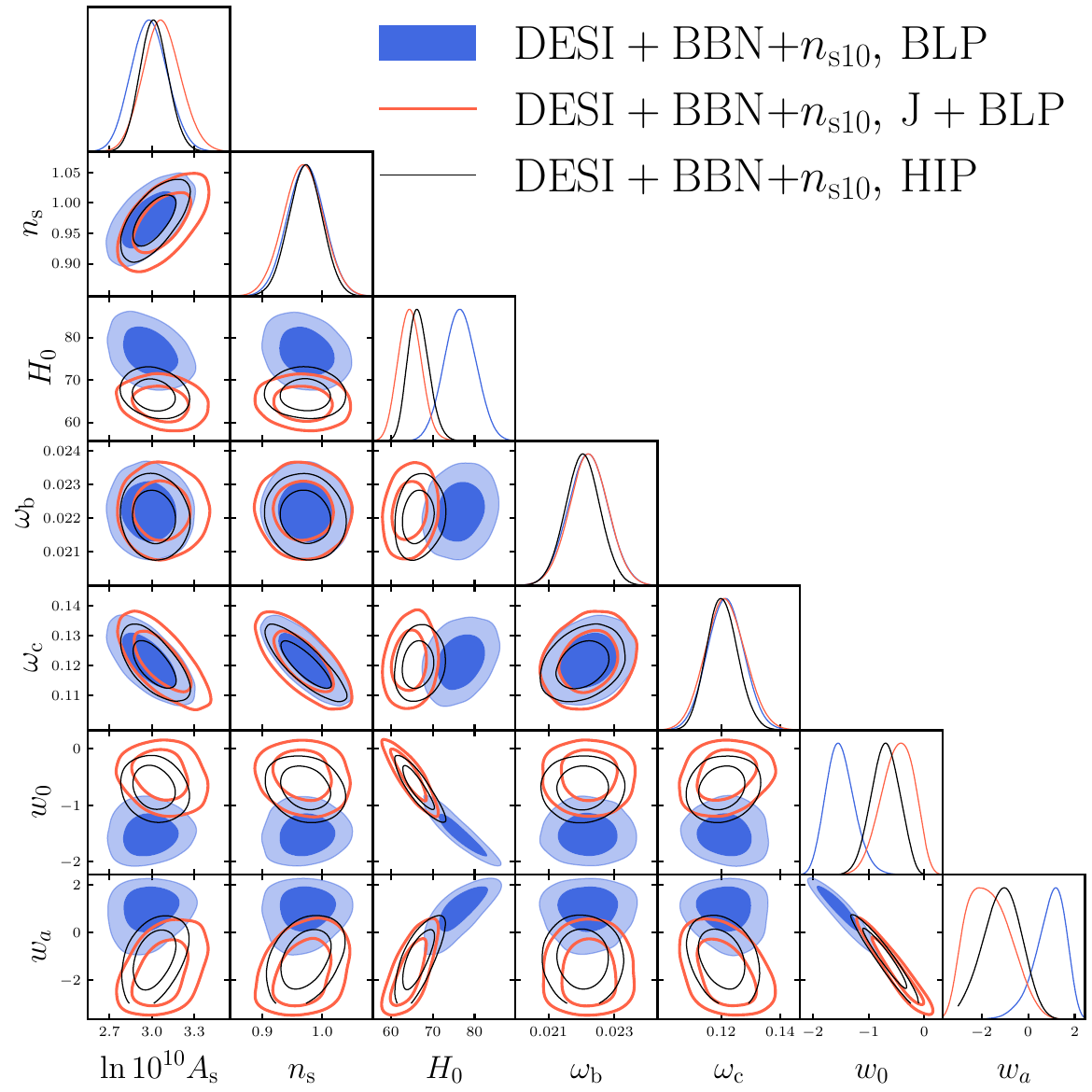}
  \caption{Marginalized one- and two-dimensional posterior distributions for cosmological parameters in the $w_{0}w_{a}$CDM model under the $\textsc{DESI}+\textsc{BBN}+n_{\mathrm{s},10}$ dataset combination. We compare three analysis frameworks: baseline Gaussian priors (BLP, blue filled contours), hybrid Jeffreys+baseline-Gaussian (J+BLP, red contours), and HOD-informed priors (HIP, black contours). The BLP analysis exhibits strong prior-volume projection effects, with posteriors displaced substantially from the likelihood maximum, particularly evident in $H_{0}$, $w_{0}$, and $w_{a}$. Both J+BLP and HIP effectively mitigate these projection effects, yielding closely aligned posterior centroids and similar degeneracy directions. The HIP framework produces somewhat tighter constraints owing to its more restrictive, physically motivated priors on the nuisance parameters. Despite this difference in constraint width, the overall agreement in central values and correlation structures demonstrates consistent extraction of cosmological information across methodologies once projection is properly controlled.}
  \label{fig:contours_FS_BAO}
\end{figure}

As anticipated, the BLP exhibits strong projection effects in this weakly anchored setup, whereas both BLP+J and HIP substantially reduce the displacement between the marginalized posteriors and the corresponding maximum-a-posteriori values seen in Fig.~\ref{fig:contours_FS_BAO}. The agreement between the BLP+J and HIP contours should therefore be interpreted primarily as a consistency check between two distinct mitigation strategies, rather than as a demonstration of projection control by itself. In particular, the red (BLP+J) and black (HIP) posteriors occupy similar regions of parameter space and display very similar degeneracy directions, in sharp contrast to the broader and visibly displaced blue BLP contours in the same figure. Because HIP imposes tighter HOD-calibrated priors on the EFT/bias sector, it generally yields tighter cosmological constraints than BLP+J, including for $w_0$ and $w_a$. This is especially clear for $\ln(10^{10}A_\mathrm{s})$ in the DESI-only case: without a CMB anchor, $\ln(10^{10}A_\mathrm{s})$ remains partially degenerate with the linear bias $b_1$, so the more informative HIP priors compress this direction and correspondingly tighten the inferred constraint. We stress, however, that tighter HIP contours should not automatically be interpreted as better-calibrated uncertainties: HIP inherits the assumptions used to construct the HOD-informed prior, and residual prior-weight effects or mild over-constraining cannot be excluded a priori. In that respect, previous validation tests are encouraging, including improved recovery relative to the standard EFT prior and successful application to non-HOD mock data generated with SHAM~\cite{Zhang2024HOD-informedLSS}, but broader validation across additional mock families would be needed to fully quantify the calibration of the resulting uncertainties.

Here we report the 68\% constraints on the $w_0$-$w_a$ parameters for our fiducial approach 
\begin{equation}
\label{eq:w0wa_FS_BAO}
\left.
\begin{aligned}
w_0 &= -0.48^{+0.33}_{-0.25},\\
w_a &= -1.66^{+0.75}_{-1.1}
\end{aligned}
\right\}
\quad \text{DESI+BBN+$n_{s,10}$}.
\end{equation}

\begin{figure}[htbp]         % “float” so LaTeX can move it
  \centering
  % width=.9\linewidth scales the graphic to 90 % of the text width
  \includegraphics[width=\linewidth]{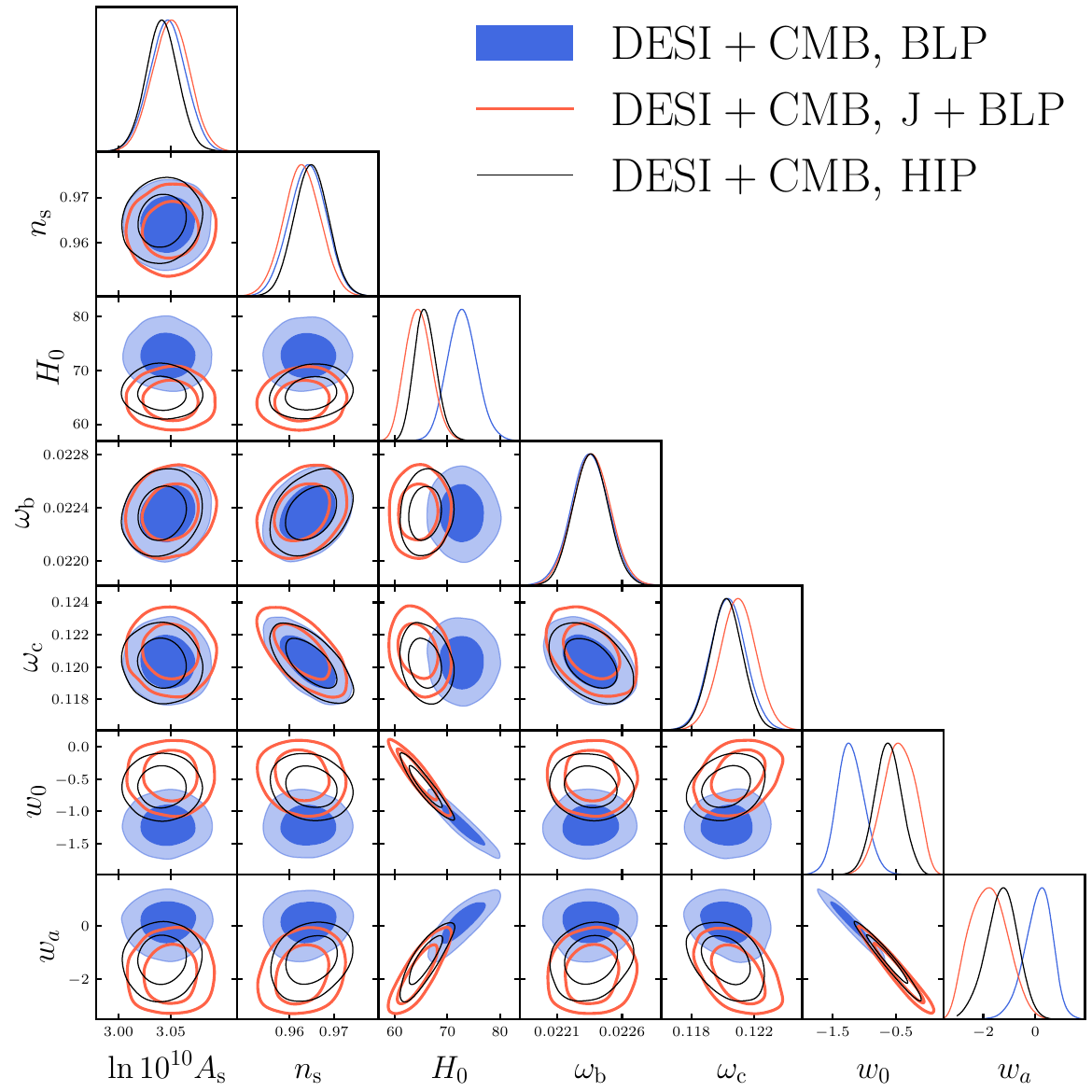}
  \caption{Marginalized one- and two-dimensional posterior distributions for cosmological parameters in the $w_{0}w_{a}$CDM model under the $\textsc{DESI}+\textsc{CMB}$ dataset combination. We compare three analysis frameworks: baseline Gaussian priors (BLP, blue filled contours), hybrid Jeffreys+baseline-Gaussian (J+BLP, red contours), and HOD-informed priors (HIP, black contours). With CMB information included, projection effects are substantially reduced compared to the $\textsc{DESI}+\textsc{BBN}+n_{\mathrm{s},10}$ case, though the BLP analysis still exhibits noticeable displacement in $H_{0}$, $w_{0}$, and $w_{a}$. Both J+BLP and HIP effectively eliminate these residual projection effects, yielding closely aligned posterior centroids and nearly identical degeneracy directions across all parameter pairs. The HIP framework continues to produce tighter constraints due to its physically motivated priors on nuisance parameters. The strong agreement between J+BLP and HIP in both central values and correlation structures confirms that both methodologies extract consistent cosmological information when projection is properly controlled, with differences attributable primarily to the restrictiveness of prior assumptions on the EFT sector rather than to inconsistencies in the underlying likelihood.}
  \label{fig:contours_FS_BAO_CMB}
\end{figure}

With the inclusion of CMB information (see Fig.~\ref{fig:contours_FS_BAO_CMB}), the BLP still exhibits large projection effects in \(w_0\), \(w_a\), and \(H_0\), whereas the CMB anchor stabilizes \(\ln(10^{10}A_\mathrm{s})\) and suppresses the amplitude–linear bias degeneracy that was prominent in the DESI-only configuration. 
Both HIP and BLP+J reduce projection-driven displacements and yield consistent centroids and degeneracy directions; the residual difference in \(\ln(10^{10}A_\mathrm{s})\) between the two essentially vanishes once CMB is included, and HIP delivers somewhat tighter contours overall due to its more informative, HOD-calibrated nuisance priors. For our approach, the 68\% confidence regions for the $w_0$-$w_a$ parameters are

\begin{equation}
\label{eq:w0wa_DESI_CMB}
\left.
\begin{aligned}
w_0 &= -0.46^{+0.28}_{-0.23}\\
w_a &= -1.78^{+0.66}_{-0.80}
\end{aligned}
\right\}
\quad \text{DESI+CMB.}
\end{equation}

\begin{figure}[htbp]         % “float” so LaTeX can move it
  \centering
  % width=.9\linewidth scales the graphic to 90 % of the text width
  \includegraphics[width=\linewidth]{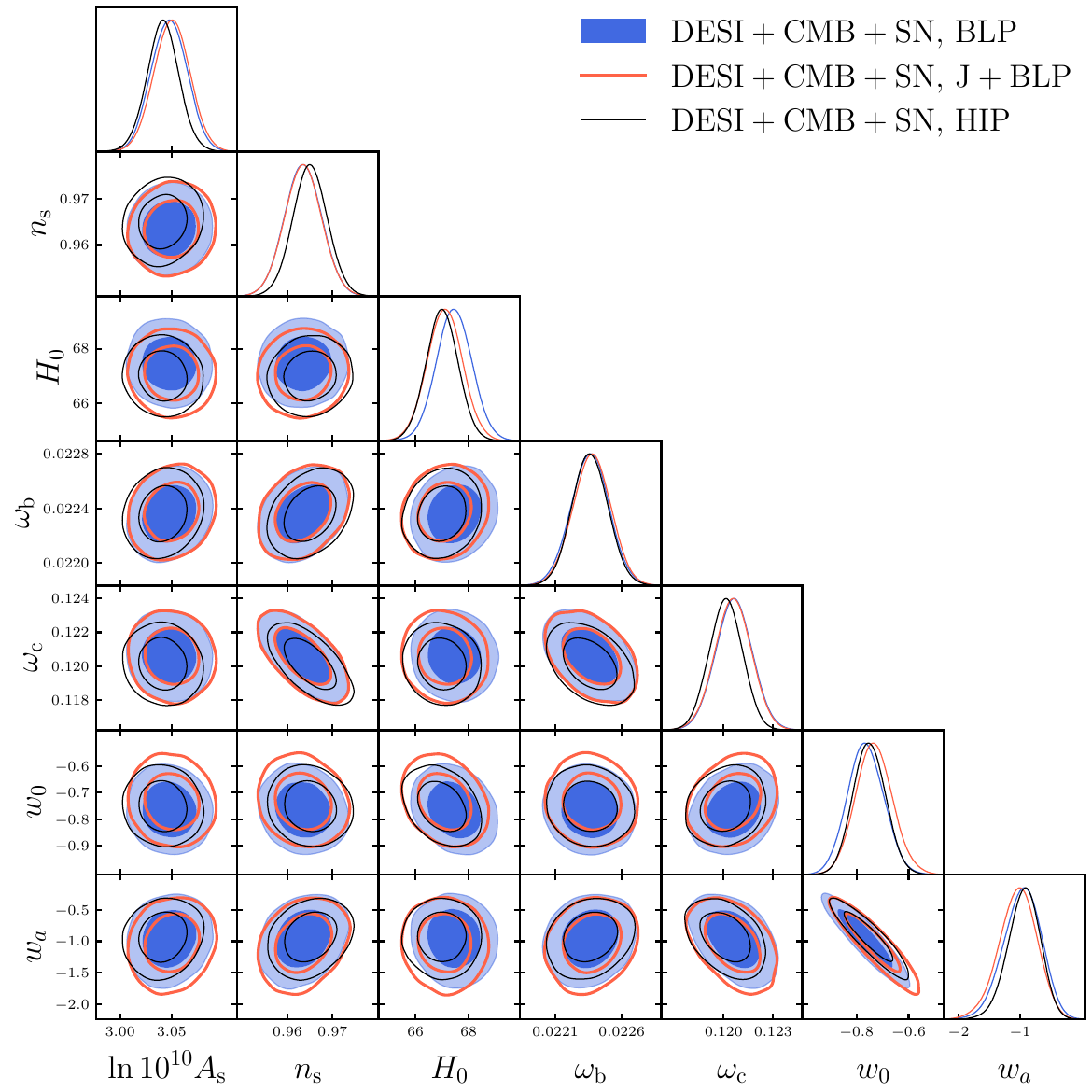}
  \caption{Marginalized one- and two-dimensional posterior distributions for cosmological parameters in the $w_{0}w_{a}$CDM model under the $\textsc{DESI}+\textsc{CMB}+\textsc{SN}$ dataset combination. We compare three analysis frameworks: baseline Gaussian priors (BLP, blue filled contours), hybrid Jeffreys+baseline-Gaussian (J+BLP, red contours), and HOD-informed priors (HIP, black contours). With Type Ia supernovae included, the dataset provides strong external constraints on the dark energy sector, effectively anchoring $w_{0}$ and $w_{a}$ and rendering projection effects negligible even in the baseline Gaussian analysis. All three methodologies now yield nearly identical posterior distributions, with overlapping contours across all parameter pairs and consistent central values. This convergence demonstrates that once sufficient external information is incorporated, the choice of projection-mitigation strategy becomes immaterial, and all approaches extract the same underlying cosmological information from the likelihood. The agreement across frameworks in this data-rich regime validates the effectiveness of the J+BLP and HIP mitigation strategies deployed in the more weakly constrained $\textsc{DESI}+\textsc{BBN}+n_{\mathrm{s},10}$ and $\textsc{DESI}+\textsc{CMB}$ cases.}
  \label{fig:contours_FS_BAO_CMB_SN}
\end{figure}

With supernovae included (see Fig.~\ref{fig:contours_FS_BAO_CMB_SN}), projection effects are already strongly suppressed by the luminosity–distance anchor, and the three pipelines—BLP, BLP+J, and HIP—show very good agreement in both centroids and degeneracy orientations, with HIP yielding the tightest constraints. The remaining differences across methods are subdominant relative to the statistical errors in this data combination, and the DE sector is consistently and sharply constrained. The 68\% posterior constraints on $w_0$ and $w_a$ for our baseline in this final scenario are

\begin{equation}
\label{eq:w0wa_DESI_CMB_SN}
\left.
\begin{aligned}
w_0 &= -0.732 \pm 0.066\\
w_a &= -1.04^{+0.32}_{-0.29}
\end{aligned}
\right\}
\quad \text{DESI+CMB+SN}
\end{equation}

\begin{figure}[htbp]         % “float” so LaTeX can move it
  \centering
  % width=.9\linewidth scales the graphic to 90 % of the text width
  \includegraphics[width=0.8\linewidth]{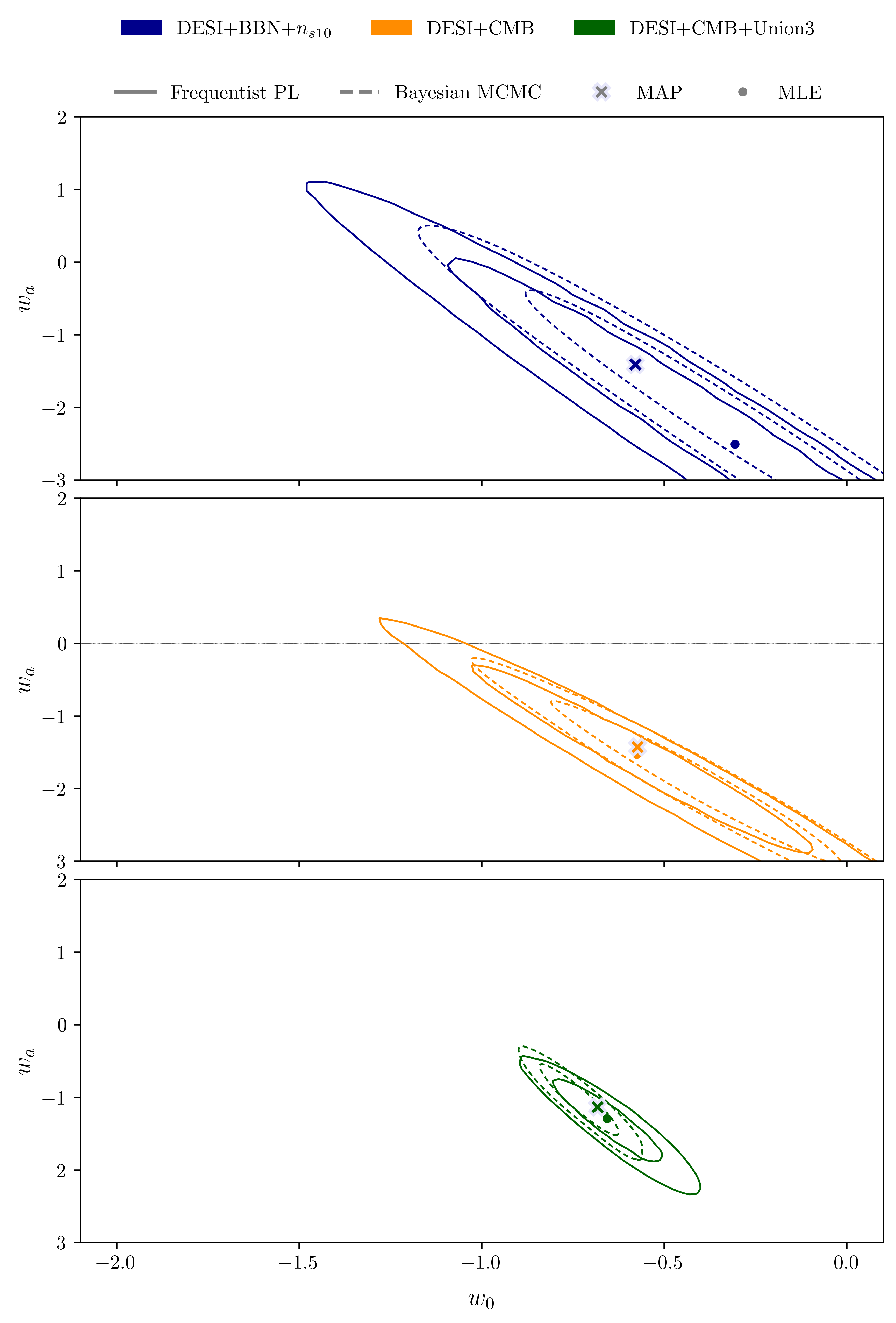}
  \caption{Comparison of frequentist profile likelihood confidence regions (solid lines) and Bayesian marginalized credible regions (dashed lines) in the $w_{0}$-$w_{a}$ plane for three dataset combinations: $\textsc{DESI}+\textsc{BBN}+n_{\mathrm{s},10}$ (blue), $\textsc{DESI}+\textsc{CMB}$ (orange), and $\textsc{DESI}+\textsc{CMB}+\textsc{Union3}$ (green). Both 68\% and 95\% regions are shown, with the MLE (gray dot) and MAP (cross) indicated. The Bayesian analysis employs the hybrid Jeffreys+baseline-Gaussian approach. Across all datasets, both frameworks exhibit excellent agreement in central values, degeneracy directions, and contour shapes, demonstrating that once projection effects are mitigated, frequentist and Bayesian approaches extract consistent cosmological information from the DESI full-shape likelihood.}
  \label{fig:2D_w0_wa}
\end{figure}

Finally, Fig.~\ref{fig:2D_w0_wa} allows a direct visual comparison between the Bayesian credible regions from our analysis and the frequentist profile-likelihood confidence contours from~\cite{desi-frequentist} for the three data combinations considered. Such a comparison should be interpreted with care, since the plotted regions do not have the same statistical meaning: Bayesian credible regions quantify posterior probability conditioned on the model, data, and priors, whereas frequentist confidence regions are constructed to satisfy coverage properties under repeated sampling, typically under asymptotic assumptions such as Wilks' theorem~\cite{Herold2024ProfileEnergy}. For this reason, differences in contour size or in the precise location of their centroids should not be over-interpreted as evidence for or against consistency between the two approaches. Nevertheless, the three panels admit a similar qualitative interpretation of the $w_0$--$w_a$ constraints, and it is notable that the MAP point in our Bayesian analysis changes only weakly across the cases shown. We therefore use Fig.~\ref{fig:2D_w0_wa} primarily as a visual cross-check between complementary inference frameworks, rather than as a demonstration of one-to-one quantitative agreement between the corresponding regions.

%\subsection{EFT parameter comparison}
\section{Conclusions}
\label{sec:conclusions}

This work addresses the prior-volume projection that arises in DESI DR1 full-shape clustering analyses when weakly constrained EFTofLSS nuisance parameters correlate with cosmology and bias marginal posteriors away from the posterior maximum \citep{DESICollaboration2024DESIMeasurements, Carrasco2012TheStructures, Ivanov2022EffectiveStructure, Simon2023ConsistencySpectrum}.
We reanalyze DESI DR1 multipoles with identical data vectors, $k$ ranges, modeling, window treatment, and covariance as the baseline pipeline, and study two complementary mitigation strategies: the orthogonalization procedure of~\cite{Paradiso2024ReducingStructure}, implemented here with a neural-network regressor in place of the original GAM-based mapping, and a fully reparameterization-invariant Jeffreys prior applied to all EFT coefficients, extending previous Jeffreys-prior implementations that were restricted to subsets of the nuisance sector~\citep{jeffreys, Donald-McCann:2023kpx, Zhao:2023ebp}, optionally combined with the original Gaussian envelope on the EFT sector to conservatively bound extreme directions~\citep{DESICollaboration2024DESIMeasurements, DESICollaboration2024DESIQuasarsb}.

Orthogonalization learns and subtracts the conditional expectation $f(\mathbf{C})\approx \mathbb{E}[\mathbf{N}\mid \mathbf{C}]$ via a neural network so that $\mathbf{N}'=\mathbf{N}-f(\mathbf{C})$ is approximately uncorrelated with cosmology, thereby reducing sensitivity to nuisance prior volume while preserving the DESI baseline modeling and data choices \citep{Paradiso2024ReducingStructure, DESICollaboration2024DESIMeasurements}.
The Jeffreys prior multiplies the likelihood by the Fisher matrix, which is evaluated at each sampling step using closed-form Jacobians for the power-spectrum multipoles, thus enforcing a coordinate-invariant measure that suppresses projection without importing HOD assumptions. Unlike the orthogonalization approach, which alleviates projection effects only at the level of posterior expectation values by learning the mean nuisance parameters conditional on fixed cosmological values, the Jeffreys prior operates directly at the likelihood level for each point in parameter space. This enables a more subtle, dynamic reparameterization that depends on all parameters simultaneously rather than fixing the transformation once the cosmological parameters are specified. A hybrid Jeffreys+baseline-Gaussian variant tempers overly broad directions while retaining the volume correction \citep{jeffreys, Donald-McCann:2023kpx, Zhao:2023ebp}.

Under the $\textsc{DESI}+\textsc{BBN}+\,n_{\mathrm{s}}$ combination, the BLP exhibits multi-$\sigma$ projection in $H_{0}$ and $(w_{0},w_{a})$, while orthogonalization partially recenters these posteriors and the full Jeffreys prior brings the MAP comfortably inside the $68\%$ regions for all three, identifying $H_{0}$, $w_{0}$, and $w_{a}$ as the parameters most severely impacted and most cleanly cured by the Jeffreys weighting \citep{DESICollaboration2024DESIMeasurements, DESICollaboration2024DESIQuasarsb, Simon2023ConsistencySpectrum}.
A residual over-broad tail in $\omega_{\mathrm{c}}$ under the pure Jeffreys prior is eliminated by reinstating the baseline EFT Gaussian widths, motivating the Jeffreys+baseline-prior prescription as the fiducial configuration for the remainder of the analysis \citep{jeffreys, DESICollaboration2024DESIMeasurements}.

With CMB information added, both Jeffreys variants enclose the MAP for $H_{0}$, $w_{0}$, and $w_{a}$, and once Type Ia supernovae are included, all the prior choices considered converge, indicating that projection effects are significantly reduced in these richer data combinations \citep{Collaboration2020PlanckParameters, DESICollaboration2024DESIMeasurements}.

Beyond the orthogonalization and Jeffreys strategies, we compare our fiducial Jeffreys+baseline-Gaussian approach against two complementary frameworks: the HOD-informed prior and the frequentist profile likelihood.
The HIP framework imposes physically motivated, tighter priors on EFT and bias parameters calibrated from HOD simulations~\cite{Zhang2024HOD-informedLSS, DESI:2025wzd}, thereby constraining the nuisance sector through external astrophysical information rather than geometric volume control.
The frequentist profile likelihood, by contrast, optimizes over nuisance parameters at each cosmological configuration to construct confidence regions with long-run coverage guarantees under asymptotic assumptions~\cite{desi-frequentist}.
Across all three dataset combinations examined—$\textsc{DESI}+\textsc{BBN}+\,n_{\mathrm{s}}$, $\textsc{DESI}+\textsc{CMB}$, and $\textsc{DESI}+\textsc{CMB}+\textsc{SN}$—these distinct methodologies yield remarkably consistent results: the centers of marginalized posteriors and confidence intervals align closely, with similar degeneracy directions in the $w_{0}$-$w_{a}$ plane.
While the HIP approach produces somewhat tighter constraints owing to its more restrictive prior assumptions on the nuisance sector, the qualitative agreement in both central values and credible region orientations demonstrates that the underlying likelihood information is being extracted coherently by all three strategies once projection effects are properly controlled.
This cross-framework consistency validates that the Jeffreys+BLP, HIP, and profile likelihood approaches can all be reliably applied to DESI full-shape data, each offering complementary perspectives—Bayesian with volume correction, Bayesian with astrophysical anchoring, and frequentist with likelihood profiling—that converge on compatible cosmological inferences.

The robustness of these results across methodological choices opens promising avenues for ongoing and future large-scale structure analyses.
The systematic agreement among debiasing strategies suggests that similar techniques can be deployed with confidence in forthcoming DESI and Euclid analyses, where increased statistical power will further test the stability of these mitigation approaches.

By establishing that multiple independent frameworks—geometric volume correction, astrophysical calibration, and frequentist profiling—yield consistent cosmological constraints from DESI DR1, this work provides a methodological foundation for robust late-time expansion and growth inference in the era of precision large-scale structure cosmology.

\appendix
\begin{figure}
    \centering
    \includegraphics[width=\linewidth]{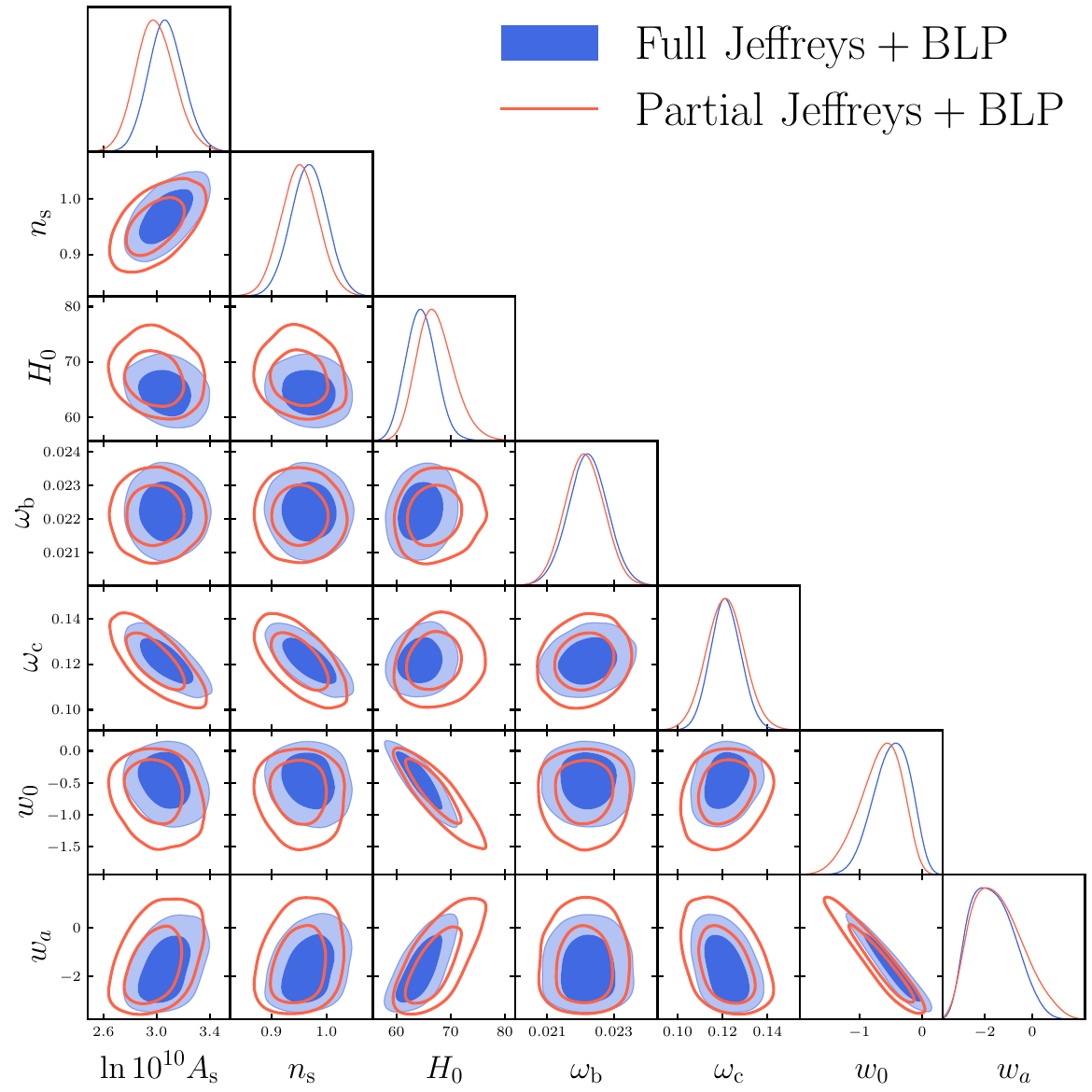}
    \caption{Triangular plot showing the posterior for the DESI+BBN+$n_{\mathrm{s},10}$ dataset, when using the full (blue) and the partial (red) Jeffreys prior.}
    \label{fig:partial}
\end{figure}

\section{The impact of the full Jeffreys prior}
\label{app:full}

In this appendix, we examine the differences between applying the Jeffreys prior to all EFTofLSS parameters (full Jeffreys) versus restricting it to only the linear parameters (partial Jeffreys).

The partial Jeffreys approach has been commonly adopted in the literature primarily for computational convenience. When employing analytical marginalization techniques, incorporating a Jeffreys prior on the linear parameters is straightforward and computationally efficient, as there is no need to explicitely evaluate the Fisher matrix. This computational advantage has made the partial Jeffreys a pragmatic choice for many analyses.

In this work, however, we extend beyond this limitation by implementing the full Jeffreys prior for all nuisance parameters. As described in Section~\ref{sec:jeffreys_prior}, we achieve this by computing the determinant of the Fisher information matrix on-the-fly at each MCMC step using analytical Jacobians. This approach allows us to flexibly choose whether to apply the Jeffreys weighting to all EFTofLSS parameters or only a subset, without compromising computational efficiency.

Figure~\ref{fig:partial} presents a direct comparison between the full and partial Jeffreys implementations for the DESI~+~BBN~+~$n_\mathrm{s}$ dataset. The two approaches yield broadly consistent posterior contours across most cosmological parameters, demonstrating that the partial Jeffreys successfully captures the dominant projection effects. However, small differences emerge in specific parameters: the marginalized posteriors for $H_0$ and $\ln(10^{10}A_\mathrm{s})$ show small but discernible shifts in their means between the two implementations.

These results indicate that while the partial Jeffreys prior effectively mitigates the bulk of prior-volume projection effects, the full Jeffreys prior provides additional refinement by addressing residual projections from the non-linear nuisance sector. The improved centering of posteriors under the full Jeffreys justifies the additional computational effort required for its implementation, particularly in weakly constrained data configurations where projection effects are most severe.

\section{Test on synthetic dataset}
\label{app:test}

We further test our fiducial prior prescription, namely the hybrid Jeffreys + baseline prior, using a synthetic data realization generated with the \texttt{velocileptors} model. The synthetic data vector includes both the BAO and full-shape contributions, and is analyzed using the same covariance matrix adopted in the main analysis.

\begin{figure}
    \centering
    \includegraphics[width=\linewidth]{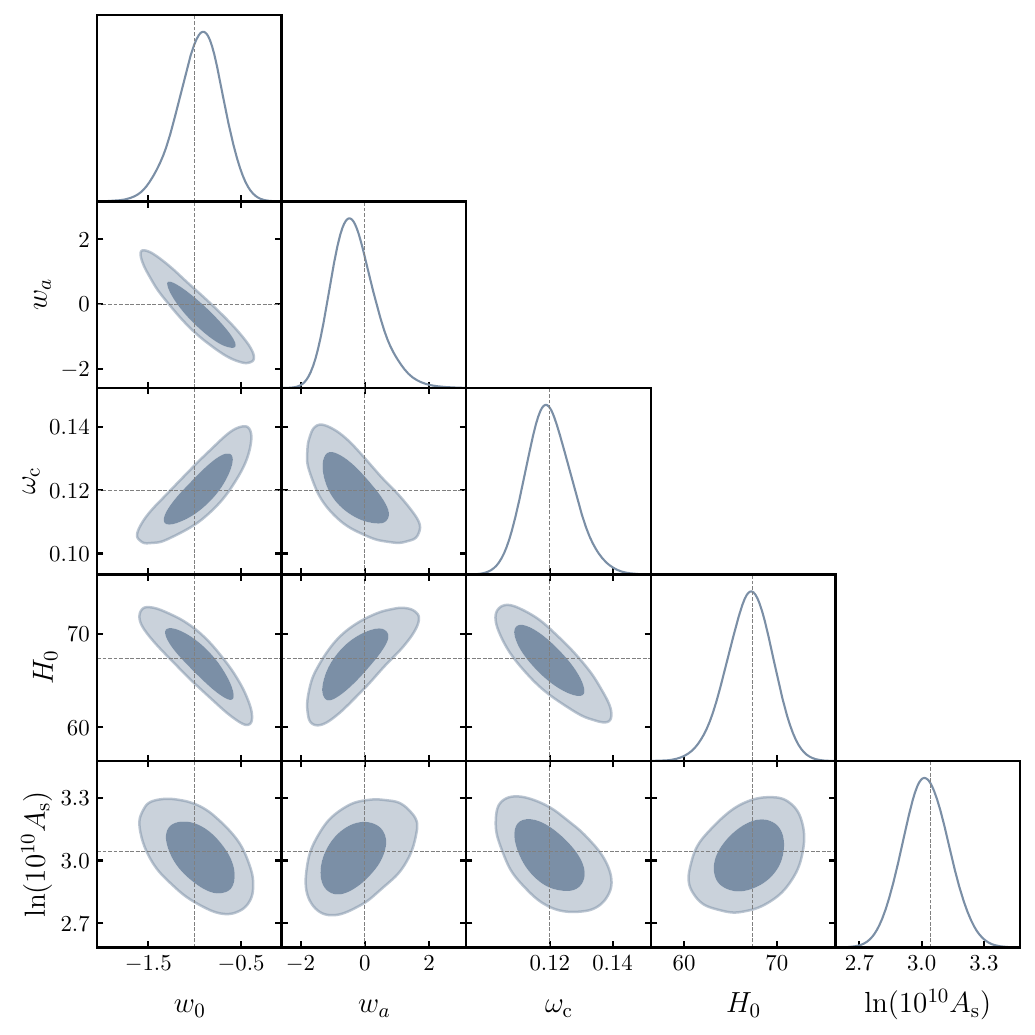}
    \caption{Triangular plot showing posterior contours for a synthetic DESI+BBN+$n_{\mathrm{s}10}$ data set, obtained using the Jeffreys+Baseline prior. The test shows recovery of input parameters, represented by dashed lines, in a controlled setup.}
    \label{fig:test}
\end{figure}

We then fit this synthetic data set in the $w_0$--$w_a$CDM model to assess whether our pipeline can accurately recover the input cosmology. As shown in Fig.~\ref{fig:test}, the posterior obtained with the Jeffreys + BLP recovers the input cosmological parameters (dashed lines), indicating that the fiducial prescription successfully suppresses the projection effects seen with the baseline prior alone. This test therefore provides an end-to-end consistency check of our analysis setup, showing that the combination of the hybrid prior, the full-shape + BAO data vector, and the adopted covariance yields a marginalized posterior that reliably recovers the underlying cosmological model.

\section{Data Availability}
The data used in this analysis are public along with Data Release 1 (details in \url{https://data.desi.lbl.gov/doc/releases/}). 
The data points corresponding to the figures from this paper is available at \url{https://doi.org/10.5281/zenodo.15178357}.

%%%%%%%%%%%%%%%%% AFFILIATIONS %%%%%%%%%%%%%%%%%%%%%
\section{Author Affiliations}
\label{sec:affiliations}
\noindent \hangindent=.5cm $^{1}${Perimeter Institute for Theoretical Physics, 31 Caroline St. North, Waterloo, ON N2L 2Y5, Canada}

\noindent \hangindent=.5cm $^{2}${Department of Physics and Astronomy, University of Waterloo, 200 University Ave W, Waterloo, ON N2L 3G1, Canada}

\noindent \hangindent=.5cm $^{3}${Waterloo Centre for Astrophysics, University of Waterloo, 200 University Ave W, Waterloo, ON N2L 3G1, Canada}

\noindent \hangindent=.5cm $^{4}${Istituto Nazionale di Astrofisica - Osservatorio di Astrofisica e Scienza dello Spazio di Bologna, via Gobetti 101, I-40129 Bologna, Italy}

\noindent \hangindent=.5cm $^{5}${Istituto Nazionale di Fisica Nucleare, Sezione di Bologna, viale Berti Pichat 6/2, I-40127 Bologna, Italy}

\noindent \hangindent=.5cm $^{6}${Department of Statistics and Actuarial Sciences, University of Waterloo, Waterloo, ON N2L 3G1, Canada}

\noindent \hangindent=.5cm $^{7}${Dipartimento di Scienze Matematiche, Fisiche e Informatiche, Università di Parma, Parco Area delle Scienze, I-43124, Parma, Italy}

\noindent \hangindent=.5cm $^{8}${INFN Sezione Milano-Bicocca, Gruppo Collegato di Parma, I-43124, Parma, Italy}

\noindent \hangindent=.5cm $^{9}${University of California, Berkeley, 110 Sproul Hall \#5800 Berkeley, CA 94720, USA}

\noindent \hangindent=.5cm $^{10}${Lawrence Berkeley National Laboratory, 1 Cyclotron Road, Berkeley, CA 94720, USA}

\noindent \hangindent=.5cm $^{11}${Department of Physics, Boston University, 590 Commonwealth Avenue, Boston, MA 02215 USA}

\noindent \hangindent=.5cm $^{12}${Dipartimento di Fisica ``Aldo Pontremoli'', Universit\`a degli Studi di Milano, Via Celoria 16, I-20133 Milano, Italy}

\noindent \hangindent=.5cm $^{13}${INAF-Osservatorio Astronomico di Brera, Via Brera 28, 20122 Milano, Italy}

\noindent \hangindent=.5cm $^{14}${Department of Physics \& Astronomy, University College London, Gower Street, London, WC1E 6BT, UK}

\noindent \hangindent=.5cm $^{15}${Institut d'Estudis Espacials de Catalunya (IEEC), c/ Esteve Terradas 1, Edifici RDIT, Campus PMT-UPC, 08860 Castelldefels, Spain}

\noindent \hangindent=.5cm $^{16}${Institute of Space Sciences, ICE-CSIC, Campus UAB, Carrer de Can Magrans s/n, 08913 Bellaterra, Barcelona, Spain}

\noindent \hangindent=.5cm $^{17}${Instituto de F\'{\i}sica, Universidad Nacional Aut\'{o}noma de M\'{e}xico,  Circuito de la Investigaci\'{o}n Cient\'{\i}fica, Ciudad Universitaria, Cd. de M\'{e}xico  C.~P.~04510,  M\'{e}xico}

\noindent \hangindent=.5cm $^{18}${Department of Astronomy \& Astrophysics, University of Toronto, Toronto, ON M5S 3H4, Canada}

\noindent \hangindent=.5cm $^{19}${Department of Physics \& Astronomy and Pittsburgh Particle Physics, Astrophysics, and Cosmology Center (PITT PACC), University of Pittsburgh, 3941 O'Hara Street, Pittsburgh, PA 15260, USA}

\noindent \hangindent=.5cm $^{20}${Instituci\'{o} Catalana de Recerca i Estudis Avan\c{c}ats, Passeig de Llu\'{\i}s Companys, 23, 08010 Barcelona, Spain}

\noindent \hangindent=.5cm $^{21}${Institut de F\'{i}sica d’Altes Energies (IFAE), The Barcelona Institute of Science and Technology, Edifici Cn, Campus UAB, 08193, Bellaterra (Barcelona), Spain}

\noindent \hangindent=.5cm $^{22}${Departamento de F\'isica, Universidad de los Andes, Cra. 1 No. 18A-10, Edificio Ip, CP 111711, Bogot\'a, Colombia}

\noindent \hangindent=.5cm $^{23}${Observatorio Astron\'omico, Universidad de los Andes, Cra. 1 No. 18A-10, Edificio H, CP 111711 Bogot\'a, Colombia}

\noindent \hangindent=.5cm $^{24}${Institute of Cosmology and Gravitation, University of Portsmouth, Dennis Sciama Building, Portsmouth, PO1 3FX, UK}

\noindent \hangindent=.5cm $^{25}${University of Virginia, Department of Astronomy, Charlottesville, VA 22904, USA}

\noindent \hangindent=.5cm $^{26}${Fermi National Accelerator Laboratory, PO Box 500, Batavia, IL 60510, USA}

\noindent \hangindent=.5cm $^{27}${Department of Astronomy, University of Texas at Austin, 2515 Speedway, TX 78712, USA}

\noindent \hangindent=.5cm $^{28}${Center for Cosmology and AstroParticle Physics, The Ohio State University, 191 West Woodruff Avenue, Columbus, OH 43210, USA}

\noindent \hangindent=.5cm $^{29}${Department of Physics, The Ohio State University, 191 West Woodruff Avenue, Columbus, OH 43210, USA}

\noindent \hangindent=.5cm $^{30}${The Ohio State University, Columbus, 43210 OH, USA}

\noindent \hangindent=.5cm $^{31}${Department of Physics, The University of Texas at Dallas, 800 W. Campbell Rd., Richardson, TX 75080, USA}

\noindent \hangindent=.5cm $^{32}${NSF NOIRLab, 950 N. Cherry Ave., Tucson, AZ 85719, USA}

\noindent \hangindent=.5cm $^{33}${Department of Physics, Southern Methodist University, 3215 Daniel Avenue, Dallas, TX 75275, USA}

\noindent \hangindent=.5cm $^{34}${Sorbonne Universit\'{e}, CNRS/IN2P3, Laboratoire de Physique Nucl\'{e}aire et de Hautes Energies (LPNHE), FR-75005 Paris, France}

\noindent \hangindent=.5cm $^{35}${Departament de F\'{i}sica, Serra H\'{u}nter, Universitat Aut\`{o}noma de Barcelona, 08193 Bellaterra (Barcelona), Spain}

\noindent \hangindent=.5cm $^{36}${Departamento de F\'{\i}sica, DCI-Campus Le\'{o}n, Universidad de Guanajuato, Loma del Bosque 103, Le\'{o}n, Guanajuato C.~P.~37150, M\'{e}xico}

\noindent \hangindent=.5cm $^{37}${Instituto Avanzado de Cosmolog\'{\i}a A.~C., San Marcos 11 - Atenas 202. Magdalena Contreras. Ciudad de M\'{e}xico C.~P.~10720, M\'{e}xico}

\noindent \hangindent=.5cm $^{38}${Instituto de Astrof\'{i}sica de Andaluc\'{i}a (CSIC), Glorieta de la Astronom\'{i}a, s/n, E-18008 Granada, Spain}

\noindent \hangindent=.5cm $^{39}${Departament de F\'isica, EEBE, Universitat Polit\`ecnica de Catalunya, c/Eduard Maristany 10, 08930 Barcelona, Spain}

\noindent \hangindent=.5cm $^{40}${Department of Physics and Astronomy, Sejong University, 209 Neungdong-ro, Gwangjin-gu, Seoul 05006, Republic of Korea}

\noindent \hangindent=.5cm $^{41}${Abastumani Astrophysical Observatory, Tbilisi, GE-0179, Georgia}

\noindent \hangindent=.5cm $^{42}${Department of Physics, Kansas State University, 116 Cardwell Hall, Manhattan, KS 66506, USA}

\noindent \hangindent=.5cm $^{43}${CIEMAT, Avenida Complutense 40, E-28040 Madrid, Spain}

\noindent \hangindent=.5cm $^{44}${Space Telescope Science Institute, 3700 San Martin Drive, Baltimore, MD 21218, USA}

\noindent \hangindent=.5cm $^{45}${University of Michigan, 500 S. State Street, Ann Arbor, MI 48109, USA}

\noindent \hangindent=.5cm $^{46}${National Astronomical Observatories, Chinese Academy of Sciences, A20 Datun Road, Chaoyang District, Beijing, 100101, P.~R.~China}

\acknowledgments
WP acknowledges the support of the Natural Sciences and Engineering Research Council of Canada (NSERC), [funding reference number RGPIN-2025-03931] and from the Canadian Space Agency.  
Research at Perimeter Institute is supported in part by the Government of Canada through the Department of Innovation, Science and Economic Development Canada and by the Province of Ontario through the Ministry of Colleges and Universities.
This research was enabled in part by support provided by Compute Ontario (computeontario.ca) and the Digital Research Alliance of Canada (alliancecan.ca).

This material is based upon work supported by the U.S. Department of Energy (DOE), Office of Science, Office of High-Energy Physics, under Contract No. DE–AC02–05CH11231, and by the National Energy Research Scientific Computing Center, a DOE Office of Science User Facility under the same contract. Additional support for DESI was provided by the U.S. National Science Foundation (NSF), Division of Astronomical Sciences under Contract No. AST-0950945 to the NSF’s National Optical-Infrared Astronomy Research Laboratory; the Science and Technology Facilities Council of the United Kingdom; the Gordon and Betty Moore Foundation; the Heising-Simons Foundation; the French Alternative Energies and Atomic Energy Commission (CEA); the National Council of Humanities, Science and Technology of Mexico (CONAHCYT); the Ministry of Science, Innovation and Universities of Spain (MICIU/AEI/10.13039/501100011033), and by the DESI Member Institutions: \url{https://www.desi.lbl.gov/collaborating-institutions}. Any opinions, findings, and conclusions or recommendations expressed in this material are those of the author(s) and do not necessarily reflect the views of the U. S. National Science Foundation, the U. S. Department of Energy, or any of the listed funding agencies.

The authors are honored to be permitted to conduct scientific research on I'oligam Du'ag (Kitt Peak), a mountain with particular significance to the Tohono O’odham Nation.

The authors acknowledge the use of the NASA astrophysics data system \url{https://ui.adsabs.harvard.edu} and the arXiv open-access repository \url{https://arxiv.org}. The software was hosted on the GitHub platform \url{https://github.com}. The manuscript was typeset using the overleaf cloud-based LaTeX editor \url{https://www.overleaf.com}.

%\paragraph{Note added.} This is also a good position for notes added after the paper has been written.

% Bibliography

%% [A] Recommended: using JHEP.bst file
\bibliographystyle{JHEP}
\bibliography{references}

\end{document}